\documentclass[preprint]{elsarticle}
\usepackage{enumitem}
\usepackage{graphicx}
\usepackage{amssymb}
\usepackage{amsmath}
\usepackage{amsthm}
\usepackage{mathrsfs}
\usepackage{multirow}
\usepackage{bbding}
\usepackage{float}
\usepackage{color}
\usepackage[latin9]{inputenc}
\providecommand{\tabularnewline}{\\}

\newcommand{\beq}{\begin{equation}}
\newcommand{\eeq}{\end{equation}}
\newcommand{\ignore}[1]{}
\newcommand{\be}{\begin{equation}} \newcommand{\ee}{\end{equation}}
\newcommand{\bea}{\begin{eqnarray}} \newcommand{\eea}{\end{eqnarray}}

\begin{document}

\title{Dipole anisotropy in sky brightness 
and source count distribution in radio NVSS data }
\author{Prabhakar Tiwari, Rahul Kothari, Abhishek Naskar,  Sharvari Nadkarni-Ghosh and Pankaj Jain}
\address{Department of Physics, Indian Institute of Technology, Kanpur - 208016, India}
\date{5 March, 2014}
\begin{abstract}
We study the dipole anisotropy in 
number counts and flux density weighted number counts {or sky brightness} in the 
  NRAO VLA Sky Survey (NVSS) data. 
The dipole anisotropy is expected due to our local motion with respect
to the CMBR rest frame. 
{We analyse data with an improved fit to the number density,
$n(S)$, as a function of the flux density $S$, which allows deviation from a 
pure power law behaviour. We also impose more stringent cuts to remove
the contribution due to clustering dipole.} 
In agreement with earlier results, we find  
that the
amplitude of anisotropy is significantly larger in comparison
to the prediction based on CMBR measurements. 
 {
The extracted speed 
is found to be roughly 3 times 
 the speed corresponding to CMBR.
 The significance of   
deviation is smaller, roughly 2 $\sigma$, 
in comparison to earlier estimates.
For the cut, $S>30$ mJy, the speed is found to be $1110\pm370$ Km/s using
the source count analysis.  
The direction of the dipole anisotropy 
is found to be approximately in agreement with CMBR. 
We find that the results are 
relatively insensitive to the lower as well as upper 
limit imposed on the flux density.} 
 Our results suggest that the Universe is intrinsically anisotropic with
the axis of anisotropy axis pointing roughly towards the {CMBR dipole direction}.
Finally we present a method which may allow an independent extraction
of the local speed and an intrinsic dipole anisotropy, provided
 a larger data set becomes available in future. 
\end{abstract}
\begin{keyword}
radio galaxies: high-redshift, galaxies: active
\end{keyword}
\maketitle
\section{Introduction}
\label{sec:Intro}

The observed dipole anisotropy in the 
Cosmic Microwave Background Radiation (CMBR)
is generally interpreted in terms of the motion of the solar system 
with respect to the CMBR rest frame. 
The corresponding speed is found to be (369$\pm$ 0.9 Km s$^{-1}$) in
the direction, $l=263.99^o\pm 0.14^o$, $b=48.26\pm 0.03^o$ in galactic 
coordinates \citep{Kogut:1993,Hinshaw:2009}. In J2000 equatorial coordinates, the direction parameters are 
$RA=167.9^o$, $DEC=-6.93^o$.
We expect that, at large distance scales,
galaxies would be distributed isotropically with respect to the CMBR rest
frame. However, due to Doppler and aberration effect, they  
would show an anisotropic distribution \citep{Ellis:1984} in the solar rest
frame. 
The dipole anisotropy in radio sources should be 
observable both in number counts and flux density weighted number 
counts (sky brightness) of sources
at high redshifts. The resulting anisotropy has been probed 
in radio surveys by many authors \citep{Baleisis:1998,Blake:2002,Crawford:2009,Singal:2011,Gibelyou:2012,Rubart:2013}.     
Using the  NRAO VLA Sky Survey (NVSS) \citep{Condon:1998},  
\cite{Blake:2002} reported a positive detection of dipole anisotropy
in the radio source count with  
the direction in reasonable agreement with CMBR. The
speed was found to be roughly 1.5 to 2 times larger in comparison
to that extracted from the CMBR dipole.  
In an independent analysis of the same data, using both number counts 
and sky brightness, 
 \cite{Singal:2011}
reported a much larger value of the 
local  velocity ($\sim1600\pm$ 400 Km s$^{-1}$), which is roughly  four 
times larger 
than the prediction from CMBR observations. The direction was found to be
consistent with the CMBR dipole. The results using 
 number counts and sky brightness agreed with one another within errors,
with the sky brightness  
based analyses yielding a larger value.    
Furthermore, \cite{Gibelyou:2012,Rubart:2013} also find a dipole anisotropy
much larger than 
the CMBR prediction.

The current situation with radio analysis is clearly puzzling. All the four 
analysis \citep{Blake:2002,Singal:2011,Gibelyou:2012,Rubart:2013} 
find direction in
agreement with the CMBR, but disagree 
on the extracted speed. Although, \cite{Blake:2002}
claim results roughly 
consistent with CMBR, the large amplitude found in 
\cite{Singal:2011} suggests a potential
violation of the cosmological principle. Ref. \cite{Gibelyou:2012} attribute
their deviation from CMBR predictions to observational bias.   
 Ref. \cite{Rubart:2013}
 attribute the 
different results obtained in literature to difference in 
the estimators used. 
In the present paper we revisit this problem in an attempt to get a consistent
result. 

 {In contrast to the above results, the Planck team finds further evidence
that CMBR dipole arises dominantly due to local motion \citep{Aghanim:2013}. 
A dipole anisotropy has also been observed 
in the diffuse x-ray background \citep{Boughn:2002}. 
In this case, both the amplitude
and direction are found to be consistent with the prediction based
on the CMBR dipole. Furthermore several authors have observed that the
local galaxy distribution shows a dipole whose direction shows reasonable
agreement with the CMBR dipole 
\citep{Strauss:1992,Erdogdu:2006,Bilicki:2011}. However the magnitude
of this clustering dipole does not show convergence even up to distances
of order 300 Mpc. It appears to us that the evidence for CMBR dipole being
a kinematic effect is quite strong. All studies of dipole in galaxy surveys
also give a direction close to CMBR. However the amplitude of the dipole
in these surveys is still uncertain. Theoretically it is possible that large
scale structures might show an anisotropy not consistent with that expected
from kinematic effects. A possible model in which this can arise is
discussed in \citep{Aluri:2012,Rath:2013,Ghosh:2014}. In this case it
is assumed that the Universe is anisotropic during the pre-inflationary
phase of its evolution. The perturbations generated during this phase
re-enter the horizon at late times and can affect structure formation,
while having negligible effect on CMBR.
Hence in this case large scale structures can show an intrinsic 
dipole pattern even if the CMBR is isotropic, up to kinematic effects. 
}

The NVSS data set does not cover the sky evenly. Further cuts imposed on the data that remove the {galactic plane and the} local structure dipole exacerbate this problem. There are various approaches to analyse data that covers the sky partially. 
 We treat the masked regions by filling them randomly by data extracted 
from the remaining pixels. This preserves
the distribution of data in the masked regions. We extract the dipole from data by making a spherical harmonic decomposition. This procedure differs
from that adopted in \citep{Blake:2002} who analyze masked sky data directly.
Singal \cite{Singal:2011} adopts a different strategy of dropping sources that are diametrically opposite to the NVSS gap so as to eliminate a dipole structure introduced by the gap.
We find the full sky analysis convenient since it allows us to probe the
dipole directly. Furthermore any bias that might be generated by random
filling can be determined by simulations and removed from the final result.

 We independently extract the
dipole anisotropy from number counts and sky brightness using  
spherical harmonic decomposition. 
We also
report the statistical significance of the extracted dipole anisotropy.
This important parameter has not been reported in any of the earlier papers
which only list the sigma value with which the extracted dipole differs
from that predicted by CMBR. The
 bias in the extracted magnitude and direction of velocity, 
 generated due to partial sky data, is computed by simulations
and subtracted from the final results. 
 {We point out that a pure power law does not provide a good to the
number density of sources, $n(S)$, as a function of the flux density, $S$.
Hence we consider an improved fit which accounts for the deviation 
of number density from a pure power law.
This provides a more reliable extraction of the local velocity from
the data.}
Furthermore, we show that in this case the kinematic dipole is different for 
number counts in comparison to flux weighted number counts.
Earlier 
results indicate the presence of an intrinsic dipole anisotropy in the 
number count distribution \citep{Singal:2011}. 
 We show that, with sufficient data, the improved fit
allows an independent extraction of the local velocity and the intrinsic
dipole anisotropy, which might be present in data.  
Finally, we present a method based on flux density per 
source, which 
can be used even in the case of uneven distribution of
source counts.     
This allows an independent extraction of the local velocity. 
This method is found to be not effective for the present data set but
may prove useful when additional data becomes available.

The dipole anisotropy in the NVSS data may also get contributions 
from sources in the local supercluster, which are expected to show
a dipole with prefered direction close to the CMBR dipole. This 
contribution has to be eliminated from data. We follow two different 
procedures for this purpose. First we eliminate the known local sources using
standard catalogues.  {This procedure was also employed in 
\cite{Blake:2002}. 
For imposing this cut, we also use 
recent catalogues of points sources which have become available 
since the publication of \cite{Blake:2002}. 
Alternatively we impose an additional cut which
eliminates the supergalactic plane \cite{Singal:2011}. In this case we try
several different cuts, corresponding to supergalactic latitude, $|b'|<5^o,
10^o, 15^o$ and $20^o$.} 

We also impose a cut in order to eliminate several extended 
and bright radio sources.
In the NVSS survey these would be misidentified
 as a cluster of very large number of sources over sky regions
of size greater than a degree \citep{Blake:2002} 
and hence may introduce significant error in the extracted dipole. 
This cut was imposed in \citep{Blake:2002} 
where the authors identified a
total of 22 such regions. Such regions would introduce spurious clustering
in the data and have to be removed.  
 This cut was not imposed in subsequent analysis of the
data \citep{Singal:2011,Gibelyou:2012,Rubart:2013}. We determine the
sensitivity of the extracted dipole to this cut in order to make a 
proper comparison of results with those obtained in \citep{Blake:2002}.

The results obtained by \cite{Singal:2011} suggest the possibility that
the Universe may be intrinsically anisotropic with the preferred axis 
approximately in the direction of the CMBR dipole. There already
exist considerable evidence in favor of such a hypothesis 
\citep{Ralston:2004}.
The radio polarization offset angles with respect to the galaxy
axis show a dipole anisotropy with preferred axis closely aligned with 
CMBR dipole \citep{Jain:1998r}. 
The Cosmic Microwave Background Radiation (CMBR) quadrupole and octupole
\citep{Tegmark:2004}
as well as the two point correlations in 
the quasar optical polarizations 
\citep{Hutsemekers:1998,Hutsemekers:2000fv,Jain:2003sg} also indicate a 
preferred axis closely 
aligned with CMBR dipole axis \citep{Ralston:2004,Schwarz:2004}. 
{Another interesting effect which has been observed in CMBR is the 
hemispherical anisotropy \citep{Eriksen:2004}. This effect is generally
interpreted in terms of dipole modulation in the temperature field
\citep{Prunet:2005,Gordon:2005}. This dipole modulation effect, however,
has not been observed in \citep{Fernandez:2013,Hirata:2009} in NVSS data. 
A search for quadrupolar power anisotropy also yields a null result
\citep{Hirata:2010}. 
We point
out that this dipole modulation effect or the quadrupolar anisotropy 
is not related to the dipole
anisotropy we study in the present paper. In particular, 
the axes of the two dipoles
are completely different. }

\section{Theory}
The absolute cosmological frame of reference can be determined 
by observing the dipole anisotropy of CMBR or radio data. The assumption here 
is that there is no intrinsic dipole anisotropy and the observed dipole 
is purely due to local motion which causes Doppler and aberration effect. 
The very first attempt 
to determine the local motion began with \cite{Ellis:1984}. 
An observer moving with a velocity $\vec v$ ($v<<c$), sees the sky brighter in 
forward direction due to Doppler boosting and aberration
effect. The flux density of radio sources follows 
a power-law  dependence on frequency $\nu$, $S \propto\nu^{-\alpha}$,
with $\alpha\approx 0.75$ \citep{Ellis:1984}. 
For an observer moving with a velocity $v$,
the Doppler shift in the frequency ($\nu_{rest}$) is 
$\nu_{obs} =\nu_{rest} \delta$, where
\begin{equation}
\delta \approx 1+(v/c) \cos \theta \  , 
\label{eq:delta}
\end{equation}
at leading order. 
This leads to \citep{Ellis:1984},
\begin{equation}
S_{obs}=S_{rest} \delta^{1+\alpha}
\label{eq:Sobs}
\end{equation}
 at a fixed frequency
in observer frame. 
Furthermore, the
aberration effect changes the solid angle in the direction of motion
$d\Omega_{obs}=d\Omega_{rest} \delta^{-2}$.

\begin{figure}[h!]
\includegraphics[width=4.5in,angle=0]{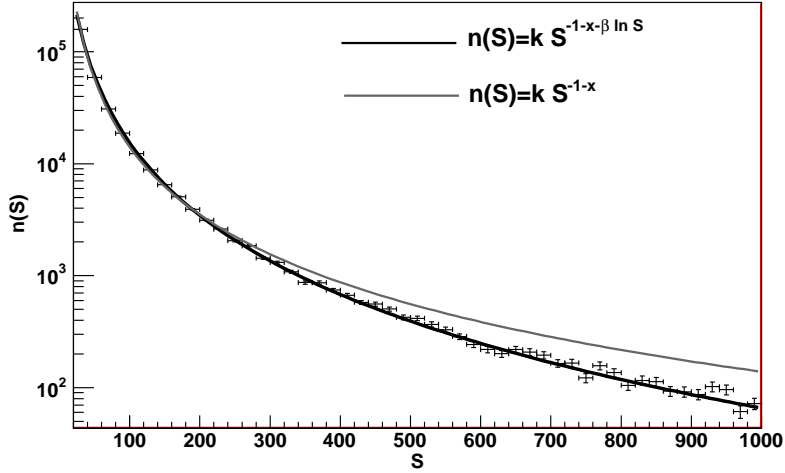}
\caption{The fits to the distribution of number counts per unit flux density,
$dN/dS$, 
based on the two functional forms, given in Eqs. \ref{eq:d2N} (upper curve) and 
\ref{eq:d2Ng} (lower curve),
over the flux density range 20 mJy $-$ 1000 mJy. The fit parameters are
(a) $x=1.006$ and (b) $x=-0.1335$, $\beta=0.1307$.  
}
\label{fig:fit}
\end{figure}

Let us denote the differential number count per unit solid angle per unit
flux density by $n(\theta,\phi,S)$, where $(\theta,\phi)$ are the polar 
angles corresponding to the direction of 
observation. Assuming isotropy, we have, in the rest frame,
\begin{equation}
 n_{rest}(\theta,\phi,S_{rest})\equiv {d^2N_{rest}\over d\Omega_{rest} dS_{rest}} =  {k x} \left({S_{rest}}\right)^{-1-x} 
\label{eq:d2N}
\end{equation}
where we have assumed a power law form for $n_{rest}(\theta,\phi,S_{rest})$
and the flux density is in units of mJy.  
Here $d^2N_{rest}$ is the number of sources in a small bin, $d\Omega_{rest} 
dS_{rest}$ in solid angle
 and flux density. 
The anisotropy due to kinematic dipole is small. Hence we may fit this 
functional form to data, ignoring this anisotropy.
The best fit to data in the flux density range, 20 mJy $-$ 1000 mJy, 
over the entire sky, is shown in Fig. \ref{fig:fit}. 
In this figure $n(S)$ denotes the distribution in Eq. \ref{eq:d2N}. 
The fit parameters are determined by using 
the ``Minuit" Minimization Package provided by the CERN
ROOT software. The program applies the $\chi^2$ minimization procedure  
on binned data. Here we use uniform
 bins of width 0.1 mJy. The values of $x$ for different cuts on 
the flux density $S$ are given in Table \ref{tb:power_para}. 
For the case of $S>20$ mJy, we find that $\chi^2= 9182$ for  7416 
degrees of freedom. For a small bin size of 0.1 mJy, we find that there
are several bins which have no sources. These are removed by the fitting
program. The fit parameters show very little dependence on the choice of  
bin size. For example, if we determine the fit parameters using only 
20 bins, we obtain $x=1.064$ for the cut
$S< 20$ mJy.  
Such a mild change has
negligible effect on our final results. 

\begin{table}[h!]
\begin{tabular}{|c|c|c|c|c|c|}
\hline
$S (mJy) >$ & 10 & 20 & 30 & 40 & 50 \tabularnewline
\hline
$x$ & $0.902$&  $1.006$ & $1.072$ & $1.123$ &  1.168\tabularnewline
\hline
\end{tabular}
\caption{The parameter $x$ corresponding to
the fit, Eq. \ref{eq:d2N},
for various cuts on flux density. 
}
\label{tb:power_para}
\end{table}

 {It can be seen from Fig. \ref{fig:fit} that a pure power law, $n\propto S^{-1-x}$ does not provide a good fit to data. A much better fit is provided by 
allowing the spectral index to depend on $S$. We discuss this improved fit 
in section \ref{sec:ImprovedFit}, where we also develop the formalism
to extract the kinematic dipole for such a modified functional form. Here
we review the standard treatment for a pure power law fit, 
first developed in \citep{Ellis:1984}.} 

Let $d^2N_{obs}$ represent the number of sources 
in the bin $d\Omega_{obs} dS_{obs}$. We have,
$$ d^2N_{obs} = d^2N_{rest}$$
where $d^2N_{rest}$ is number of sources in the corresponding bin
in rest frame. We, therefore, obtain, 
\begin{equation}
 d^2N_{obs} = d^2N_{rest} = n_{rest} d\Omega_{rest} dS_{rest}
 = {k x}\left({S_{rest}}\right)^{-1-x} d\Omega_{obs}\delta^2 dS_{rest}
\label{eq:d2N1}
\end{equation}
Substituting for $S_{rest}$, we obtain,
\begin{equation}
 d^2N_{obs} = 
  {k x}\left({S_{obs}}\right)^{-1-x} \delta^{2+x(1+\alpha)} dS_{obs}
d\Omega_{obs}
\end{equation}
Integrating over $S_{obs}$ from $S_{low}$ to $\infty$, we obtain the standard 
result
 \citep{Ellis:1984}, 
\begin{equation}
\left({ dN\over d\Omega}\right)_{obs} = 
  k\left({S_{low}}\right)^{-x} \delta^{2+x(1+\alpha)} 
 = \left({ dN\over d\Omega}\right)_{rest} \delta^{2+x(1+\alpha)}
\label{eq:d2N2}
\end{equation}
Hence Doppler boosting and aberration, at leading order
in $v/c$,
 produces a dipole anisotropy in source count, given by,
\beq
\label{eq:D_n}
\vec D_{N}(v) = [2+x(1+\alpha)](\vec v/c).
\eeq
In \cite{Ellis:1984} it has been pointed out  that to measure the dipole anisotropy
at $3\sigma$ significance level one needs to have more than $2\times10^{5}$
radio sources. However such a large catalogue was not available 
at that time. 
The authors analyzed 4C catalogue data (4844 sources), and determined the local velocity 
as $551\pm448$ Km s$^{-1}$.

We next generalize this calculation to sky brightness, $S_I$. This is 
defined as the  
number counts weighted 
by flux density, i.e., 
\begin{equation}
 d^2 S_I = Sd^2N 
\label{eq:flux_defn}
\end{equation}
We have, 
\begin{equation}
d^2S_{I,obs} =  S_{obs}d^2N_{obs} = S_{obs} d^2N_{rest} = S_{obs} n_{rest} d\Omega_{rest} dS_{rest}
\end{equation}
where we have used Eq. \ref{eq:d2N1}. This leads to, 
\begin{equation}
 d^2S_{I,obs} = 
  {k x}\left({S_{obs} }\right)^{-x} \delta^{2+x(1+\alpha)} dS_{obs}
d\Omega_{obs}
\label{eq:d2S1}
\end{equation}
The integral of this quantity over $S_{obs}$ from $S_{low}$ to $\infty$ is
divergent for $x< 1$. Hence we also impose an upper limit, $S_{up}$, 
as done in the actual data analysis. 
We obtain, after integration,
\begin{equation}
\left({ dS_I\over d\Omega}\right)_{obs} = 
  {kx\over 1-x}\left[\left({S_{up} }\right)^{1-x} - \left({S_{low} }\right)^{1-x}\right] \delta^{2+x(1+\alpha)} 
 = \left({ dS_I\over d\Omega}\right)_{rest} \delta^{2+x(1+\alpha)}
\label{eq:d2S2}
\end{equation}
The final result is same as for number counts. 
The dipole anisotropy, $\vec D_S(v)$, in $S_{I}$ is also given by the 
same formula as
$\vec D_N$,
\beq
\label{eq:D_s}
\vec D_{S}(v) = [2+x(1+\alpha) ](\vec v/c).
\eeq
This provides an independent method to compute the velocity of the solar
system with respect to the cosmological rest frame.   

Earlier analysis \citep{Singal:2011}, suggest the presence of an
intrinsic dipole anisotropy in the radio data. 
Hence let us assume the following form for the observed $D_N(obs)$. 
\beq
\label{eq:D_n2}
\vec{D}_{N}(obs)=  \vec{D}_0+[2+x(1+\alpha)](\vec{v}/c)\ ,
\eeq
where $\vec{D}_0$ represents the intrinsic dipole anisotropy in the 
number counts. 
Let us next assume that the intrinsic anisotropy in the sky brightness 
is caused entirely by the intrinsic anisotropy in the source distribution.
This assumption can be directly tested by exploring the 
anisotropy in the flux density per source, denoted by $\bar S$. 
This measure would not be affected by 
the intrinsic dipole in number counts. 
Furthermore the kinematic dipole also does not
contribute to this measure since it is same both for
number counts and flux weighted number counts. Hence, with our assumptions,
 $\bar S$ should
not show a significant signal of anisotropy,

\subsection{An improved fit} 
\label{sec:ImprovedFit}

We next generalize the calculation to a somewhat better fit to data, 
which takes into account the deviation from a pure power law behaviour. 
We assume the following form of $n(\theta,\phi,S)$,
\begin{equation}
 n_{rest}(\theta,\phi,S_{rest})\equiv {d^2N_{rest}\over d\Omega_{rest} dS_{rest}} =  {k x} \left({S_{rest}}\right)^{-1-x-\beta \ln \left(S_{rest}\right)} 
\label{eq:d2Ng}
\end{equation}
 where, as before, the flux density is in units of mJy.
This functional form significantly improves the fit to data, as can
be seen in Fig. \ref{fig:fit}.  
{We use this fit to extract the local velocity from data.}
The fit parameters are given in Table \ref{tb:improved_para} for
different cuts on the flux density.
In this case the $\chi^2/dof$ is found
6481 for  7415 degrees of freedom for the cut $S>20$ mJy.
Following the steps described above, we obtain, at 
leading order in $(v/c)\cos\theta$, 
\begin{equation}
d^2N_{obs} = kx\left({S}\right)^e   
\left( 1+ \left[2+x(1+\alpha)+ 2\beta(1+\alpha)\ln \left(S_{obs}\right)\right] {v\over c} \cos\theta\right) dS_{obs} d\Omega_{obs}
\label{eq:d2Ng1}
\end{equation}
where, 
$e={-1-x-\beta\ln (S_{obs})}$. 
Integrating over $S_{obs}$ from $S_{low}$ to $S_{up}$, we obtain,
\begin{eqnarray}
\left({ dN\over d\Omega}\right)_{obs} &=& 
k\Big\lbrace I_1 + \Big(\left[ 2+ x (1+\alpha)\right] I_1+ 2\beta (1+\alpha) I_2\Big)
{v\over c}\cos\theta\Big\rbrace\nonumber \\
 &=& \left({ dN\over d\Omega}\right)_{rest} 
\left( 1 +  a 
{v\over c}\cos\theta\right)
\label{eq:d2Ng2}
\end{eqnarray}
where
\begin{equation}
a =  2+ x (1+\alpha) + 2\beta (1+\alpha) {I_2\over I_1}\ ,
\label{eq:num_a}
\end{equation}
\begin{eqnarray}
I_1 &=& x\int_{S_{low}}^{S_{up}} dS \left({S }\right)^{-1-x-\beta \ln (S)}\\ 
I_2 &=& x\int_{S_{low}}^{S_{up}} dS \left({S }\right)^{-1-x-\beta \ln (S)}\ln (S) 
\end{eqnarray}
and 
\begin{equation}
\left({ dN\over d\Omega}\right)_{rest} = kI_1 
\end{equation}
The resulting functional form can be fitted to data, after
numerically performing the integrals, $I_1$ and $I_2$, in order to
extract the local velocity. 
\begin{table}[h!]
\begin{tabular}{|c|c|c|c|c|c|}
\hline
$S (mJy) >$ & 10 & 20 & 30 & 40 & 50 \tabularnewline
\hline
$x$ & $-0.379$&  $-0.1335$ & $-0.1831$ & $-0.1320$ &  0.0478\tabularnewline
\hline                                                                          $\beta$ & 0.1199 &  0.1307 & 0.1360   & 0.1309 &  0.1134\tabularnewline
\hline
$a$ & 3.34 & 3.55 & 3.70 & 3.82 & 3.94 \tabularnewline
\hline
$b$ & 3.82 & 3.95 & 4.04 & 4.11 & 4.16  \tabularnewline
\hline
\end{tabular}
\caption{The parameters $x$ and $\beta$ corresponding to
the improved fit, Eq. \ref{eq:d2Ng},
for various cuts on flux density.  {The values of $a$ and $b$, defined
in Eqs. \ref{eq:num_a} and \ref{eq:flux_b} respectively, are also given.} 
}
\label{tb:improved_para}
\end{table}

For the case of sky brightness we obtain, 
\begin{equation}
\left({ dS_I\over d\Omega}\right)_{obs}  
 = \left({ dS_I\over d\Omega}\right)_{rest} 
\left( 1 +  b 
{v\over c}\cos\theta\right)
\label{eq:d2Sg}
\end{equation}
where
\begin{equation}
b =  2+ x (1+\alpha) + 2\beta (1+\alpha) {I_4\over I_3}\ ,
\label{eq:flux_b}
\end{equation}
\begin{eqnarray}
I_3 &=& x\int_{S_{low}}^{S_{up}} dS \left({S }\right)^{-x-\beta \ln (S)}\\ 
I_4 &=& x\int_{S_{low}}^{S_{up}} dS \left({S }\right)^{-x-\beta \ln (S)}\ln (S) 
\end{eqnarray}
and
\begin{equation}
\left({ dS_I\over d\Omega}\right)_{rest} = kI_3 
\end{equation}
We find that in this case the kinematic dipole contributes differently
to number counts and to sky brightness. The difference, however,
is small. Nevertheless, it might be useful in future in order to 
independently extract the intrinsic and kinematic dipole from 
data. 
This may be accomplished as follows. Let the observed dipole in number 
counts and sky brightness be represented as
\begin{equation}
\vec D_N = \vec D_0 + a \vec v/c 
\label{eq:DNobs2}
\end{equation}
and 
\begin{equation}
\vec D_S = \vec D_0 + b \vec v/c\ , 
\label{eq:DSobs2}
\end{equation}
where we have again assumed that the intrinsic anisotropy in 
 sky brightness is same as that in number counts.
We obtain, 
\begin{equation}
\vec v/c = {\vec D_N-\vec D_S\over a-b}  
\label{eq:vecvbyc}
\end{equation}
Using the extracted local velocity we can determine the intrinsic dipole
from Eq. \ref{eq:DNobs2}. 
Alternatively we may determine the anisotropy in the flux density 
averaged over the number counts, $\bar S$.
The dipole in this variable
is given by,
\begin{equation}
\vec D_{\bar S} = (b-a) \vec v/c\ , 
\label{eq:DSbarobs2}
\end{equation}
where we have used 
 Eqs. \ref{eq:DNobs2} and \ref{eq:DSobs2}.  
By analysing the variable $\bar S$ we can directly extract $\vec v$, 
which can be used in Eq. \ref{eq:DNobs2} to extract $\vec D_0$.

\section{The Data}
\label{sc:data}
The  NRAO VLA Sky Survey (NVSS) \citep{Condon:1998} is a radio continuum survey covering the entire northern sky,  
$\delta>-40^{o}$. It operates at the frequency of 1.4 GHz. 
The NVSS catalogue contains 1773484 radio sources. Following \citep{Blake:2002}
we impose various 
cuts on data. We only include sources with flux density lying within the
range $S_{low}$ and $S_{up}$, where $S_{low}=10, 20, 30, 40, 50$ mJy. The data with
lower limit $S_{low}=10$ mJy is not expected to be reliable \citep{Blake:2002}. 
Hence results for this cut should be interpreted with caution. 
The upper limit,
$S_{up}$,
is generally taken to be 1000 mJy. However we also determine how the results 
change if it is changed to 900 mJy.   
We remove sources lying within the galactic latitude $|b|<15^\circ$.
{After this cut, the masked region constitutes 38\% of the sky. We test the sensitivity of our results to this cut on the galactic 
latitude.}
Furthermore, we remove the clustering dipole \citep{Blake:2002}, i.e. 
sources which might belong to the local supercluster.
  This is accomplished by removing sources within 
30 arcsec of known nearby galaxies as listed in \cite{Saunders:2000} and in the third reference 
catalogue of bright Galaxies (RC3) \citep{Vaucouleurs:1991,Corwin:1994}. 
These catalogues contain 23011 and 18351 sources respectively. 
This cut
removes a total of 13597 sources from the NVSS data set. 
{The data set obtained after removal of these sources is labelled as 
set (a). 
Besides these catalogues, 
we also use recent catalogues of local point sources 
\cite{Jarrett:2003,Huchra:2012,Bilicki:2014}  
which have become 
available since the publication of \citep{Blake:2002}. These contain a
total of 617 \cite{Jarrett:2003}, 43526 \cite{Huchra:2012} and 
928352 \cite{Bilicki:2014} sources. These catalogues lead to a removal 
of additional 64475 sources from the NVSS data set, leaving a total of 
1695412 sources. An additional cut which removes the 22 sites of bright
and extended radio sources, identified in \cite{Blake:2002} and discussed
in section \ref{sec:Intro}, leaves a 
total of 1674536 sources. These 22 sites are filled by randomly 
generated isotropic data,
following the procedure adopted in \citep{Blake:2002}. 
The data set obtained after imposing this 
more stringent cut is called set (b).  
We extract the dipole vector both for set (a) and set (b). 
} 
Using all 
of the above cuts we have about $\sim 10^5$
sources remaining in the data set. 
We determine
the sensitivity of our results to the exclusion radius
around each source. In particular we determine how the results change
if this radius is varied from $30^o$ to $45^o$.

\begin{figure}[h!]
\includegraphics[width=3.5in,angle=0]{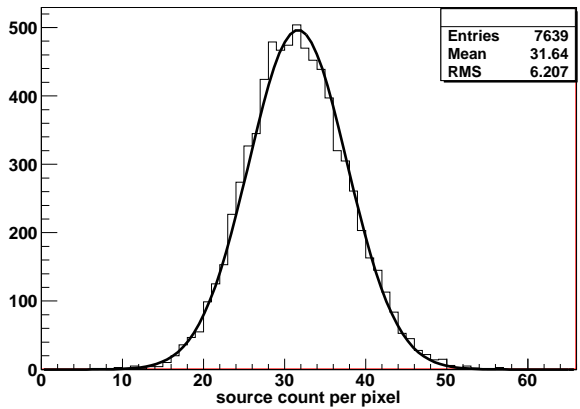}
\caption{The distribution of source counts per pixel for the cut $S>20$ mJy. 
The  Gaussian 
fit to the distribution is also shown.}
\label{fig:pix}
\end{figure}

We use HEALPix\footnote{http://healpix.jpl.nasa.gov/} to generate the angular position on the sky
with the resolution parameter $N_{side}=32$. The angular size of a pixel for  
$N_{side}=32$ is roughly
$\sim 1.8^{\circ}$. 
Using this resolution we fill the map with randomly generated data,
as described below. The 
distribution of number counts in each pixel, Fig. \ref{fig:pix}, is 
reasonably well 
described by a Gaussian. We do not find any pixel with no 
sources in the unmasked regions. 
We also determine the sensitivity of our results to the choice of
pixel size.

\section{Procedure}
\label{sc:theory}
Let $I(\theta, \phi)$, represent the number count or sky brightness 
($S_I$) as a function of 
 the polar coordinates $(\theta,\phi)$. 
Here we use J2000 equatorial system with $\phi $ equal to the
right ascension (RA) and $\theta=90^o-Dec$. 
We rewrite this field as $I(\theta, \phi) = I_{0}(1+ \Theta(\theta, \phi))$, so that $\Theta(\theta, \phi)$ represents the fluctuations in this 
field. To study the correlation of this field we expand 
$\Theta(\theta, \phi)$ in spherical harmonics, $Y_{lm}(\theta,\phi)$,  as,
\beq
\label{eq:p2a}
\Theta(\theta, \phi) = \sum_{l=1}^{\infty} \sum_{m=-l}^{+l} a_{lm}Y_{lm}(\theta,\phi),
\eeq
where, $a_{lm}$ are the harmonic coefficients. 
The power, $C_l$, in each multipole is given by, 
\beq
\label{eq:C2alm}
C_l =\frac{1}{(2l+1)} \sum_{m=-l} ^{l} |a_{lm}|^2. 
\eeq
A significant value 
of $C_l$ indicates anisotropy at a scale $\sim (\pi/l)$ radian.
In particular, $C_{1}$ represents the dipole term which 
is related to the dipole amplitude, 
$D$, as \citep{Gibelyou:2012},
\beq
\label{eq:c2d}
C_1 = \frac{4\pi}{9} D^2.
\eeq
 {In the present paper we are interested only in the dipole. Hence we do
not discuss higher multipoles in this paper.}

\begin{figure}[h!]
\includegraphics[width=3.5in,angle=0]{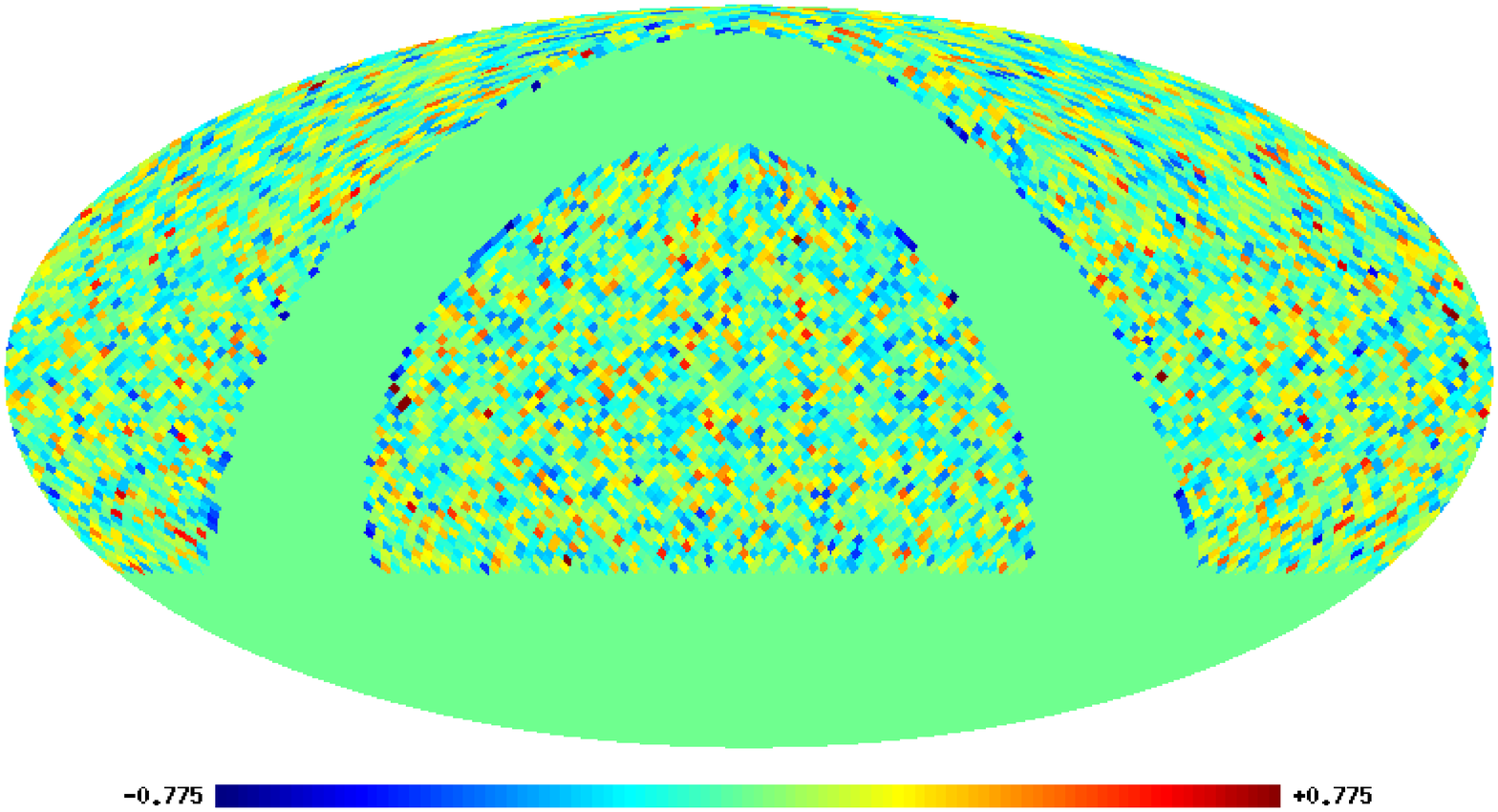}\\
\includegraphics[width=3.5in,angle=0]{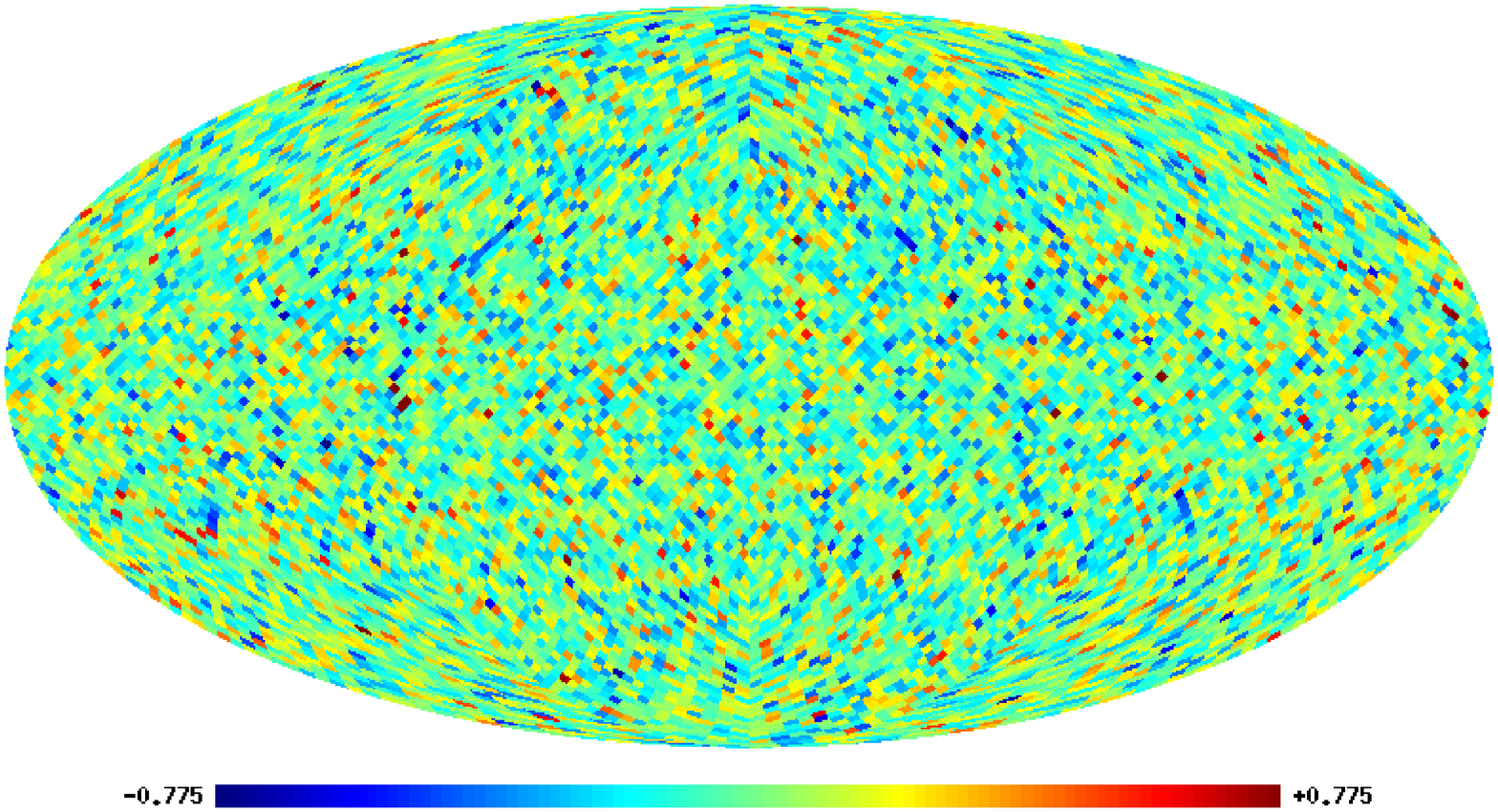}\\
\caption{ {The upper graph shows the 
NVSS source distribution map for $S>20$ mJy. The masked regions correspond to
$\delta\le -40^o$ and the galactic plane ($|b|>15^{\circ}$). The lower graph shows the source
distribution, with the masked regions filled by  
a particular realization of the randomly generated isotropic data. 
}}
\label{fig:NVSS_fill}
\end{figure}

\subsection{Analysis Method and Bias simulation}
\label{sc:anl_bias}

We do not have data for $\delta \le -40^{\circ}$ and we impose a
cut $|b|>15^\circ$ in order to eliminate the 
region around the Galactic plane. {After imposing this cut, we
have data only over  62\% of the sky area.} 
The dipole power of the real data is obtained by filling the empty 
pixels isotropically by 
 randomly generated data and computing the corresponding full sky
power. {The distribution of data in the masked region is same as real
data, as long as we ignore
 the dipole which might be present in the real data.  } 
 We generate a total of 1000 realizations of real map 
 by filling the empty pixels with different random samples. 
{The source distribution of the masked map for the cut $S>20$ mJy is shown
in Fig. \ref{fig:NVSS_fill}. In this figure we also show a full sky map, with
the masked regions filled by a particular realization of the 
randomly generated data.  
The mean dipole power, $C'_1$, and the mean 
direction parameters $(\theta',\phi')$, 
are computed from 
the 1000 realizations of the real map. The error in these parameters
is computed by simulations, as described below. }
The dipole extracted in this manner would differ
from that extracted from the full sky map, i.e. if real data in all
the pixels were available, by some constant $k$ \citep{Singal:2011}, 
which would depend 
on the mask used for data. 
Similarly 
the extracted values of the direction would also contain a bias. 
 Let $(\theta,\phi)$ represent
the true direction of the dipole and $(\theta',\phi')$
the dipole extracted after filling the masked region
by isotropic random samples. 
The bias in the angles is given by, $\Delta\theta=\theta'-\theta$ and $\Delta \phi=\phi'-\phi$. 
 We calculate the constant $k$ and the bias in direction by simulations.  
The bias corrected value of the dipole power, $C_1$, is given by,
 $C_1 =C^{'}_1/k^2$. Similarly the angle parameters of the dipole direction
are given by, $\theta = \theta'-\Delta\theta$ and $\phi=\phi'-\Delta\phi$.

\begin{figure}[h!]
\includegraphics[width=3.0in,angle=0]{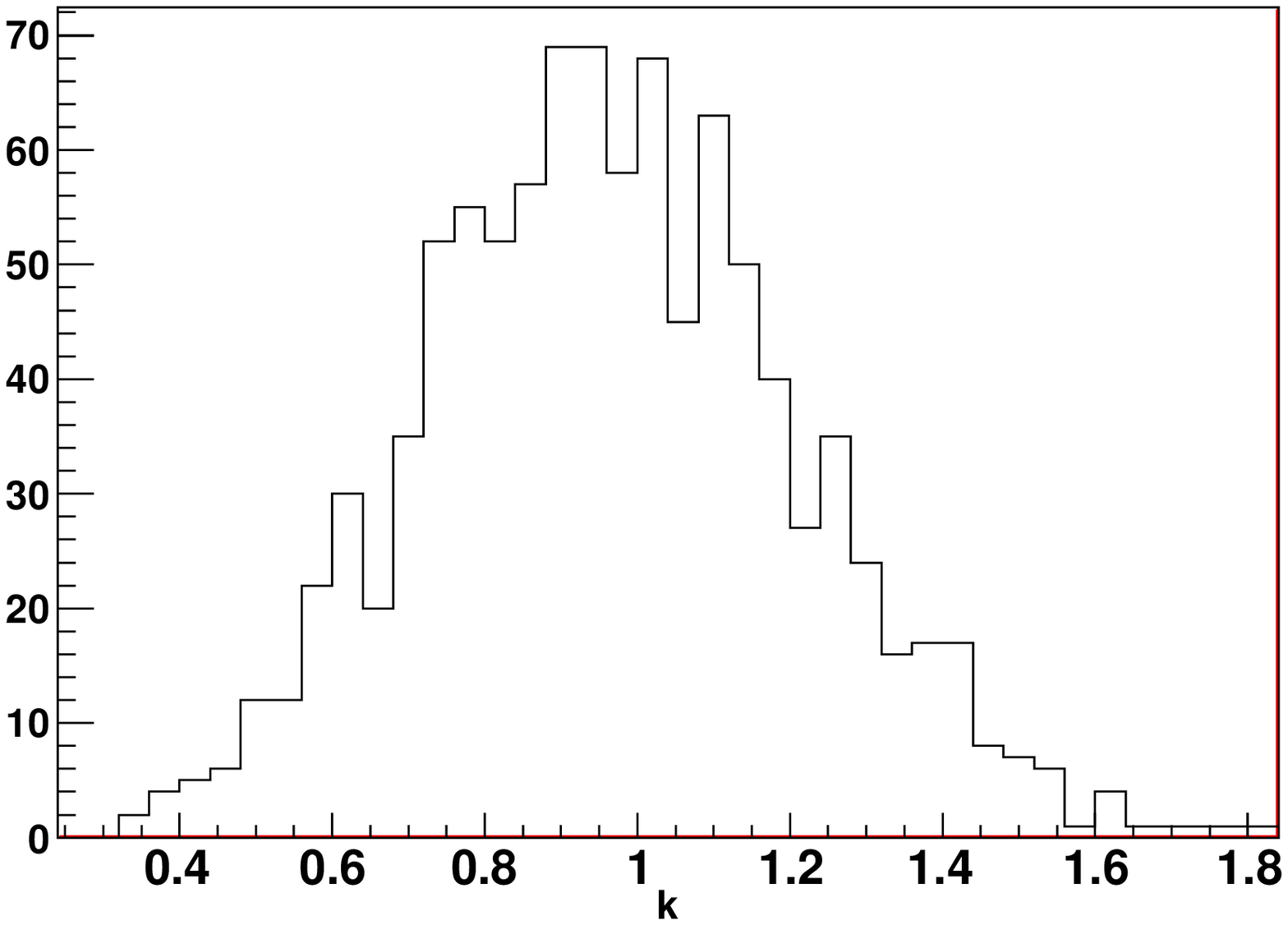}\\
\includegraphics[width=3.0in,angle=0]{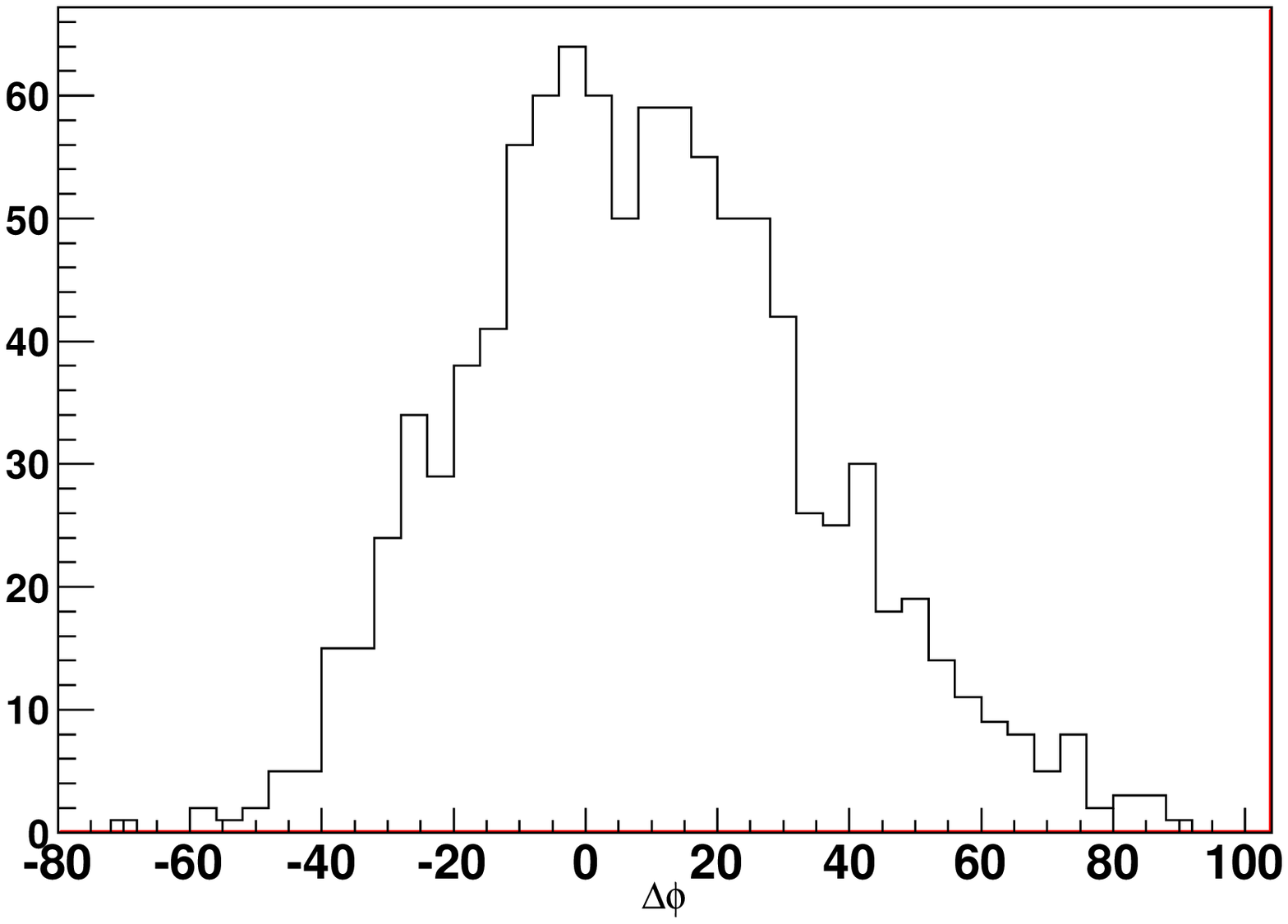} \\
\includegraphics[width=3.0in,angle=0]{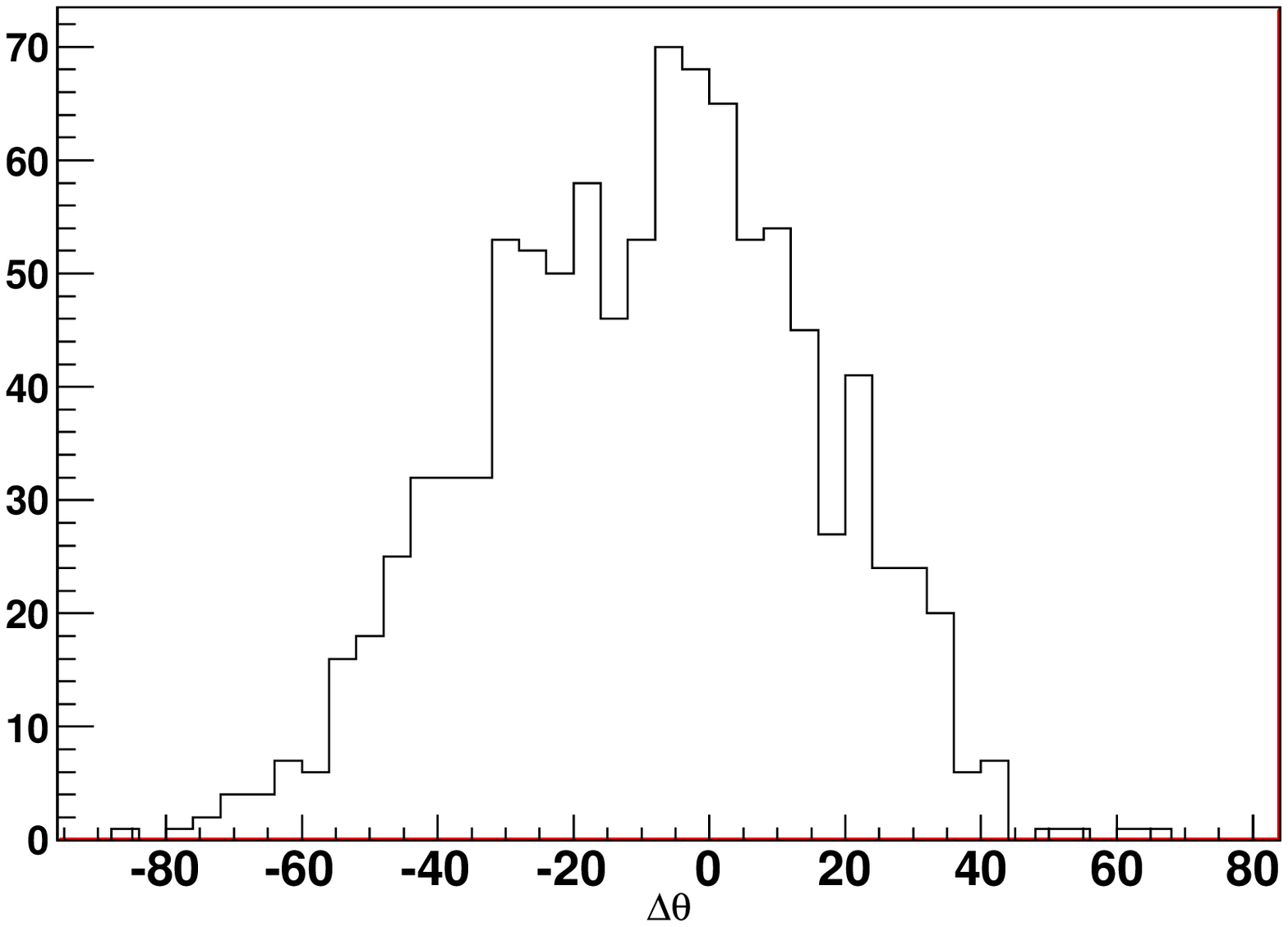}\\
\caption{
 The distribution of bias parameters, $k = \sqrt{C_1'/C_1}, \Delta \theta= \theta'-\theta$
 and $\Delta \phi = \phi'-\phi$, for sky brightness with the cut, $S>20$ mJy,
extracted by simulations. 
}
\label{fig:c1sim_PI}
\end{figure} 

The random samples for number counts are generated by using the distribution
shown in Fig. \ref{fig:pix}, which is extracted directly from data.
This is the distribution of number of sources per pixel, as explained
in section \ref{sc:data}. 
For the intensity dipole we also require the flux density of sources
in the random sample. This is 
generated by using the flux density distribution, Fig. \ref{fig:fit}, 
of real data. Using a random
sample of number counts per pixel, we randomly allocate
the values of flux density of real sources to each source in
the simulated data. 
This yields a random isotropic sample, which has same statistical properties
as the NVSS catalog.

\begin{figure}[h!]
\includegraphics[width=3.0in,angle=0]{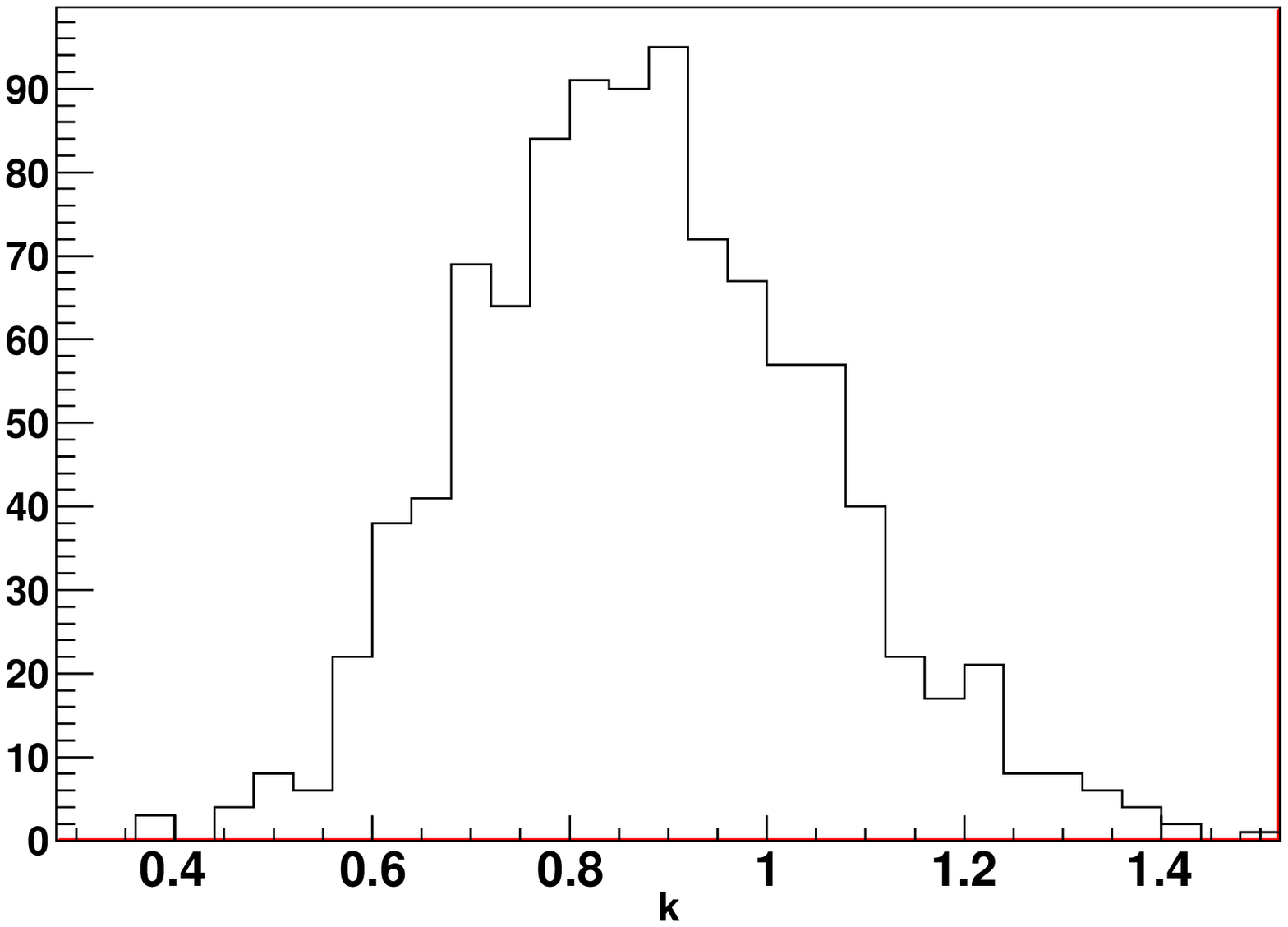}\\
\includegraphics[width=3.0in,angle=0]{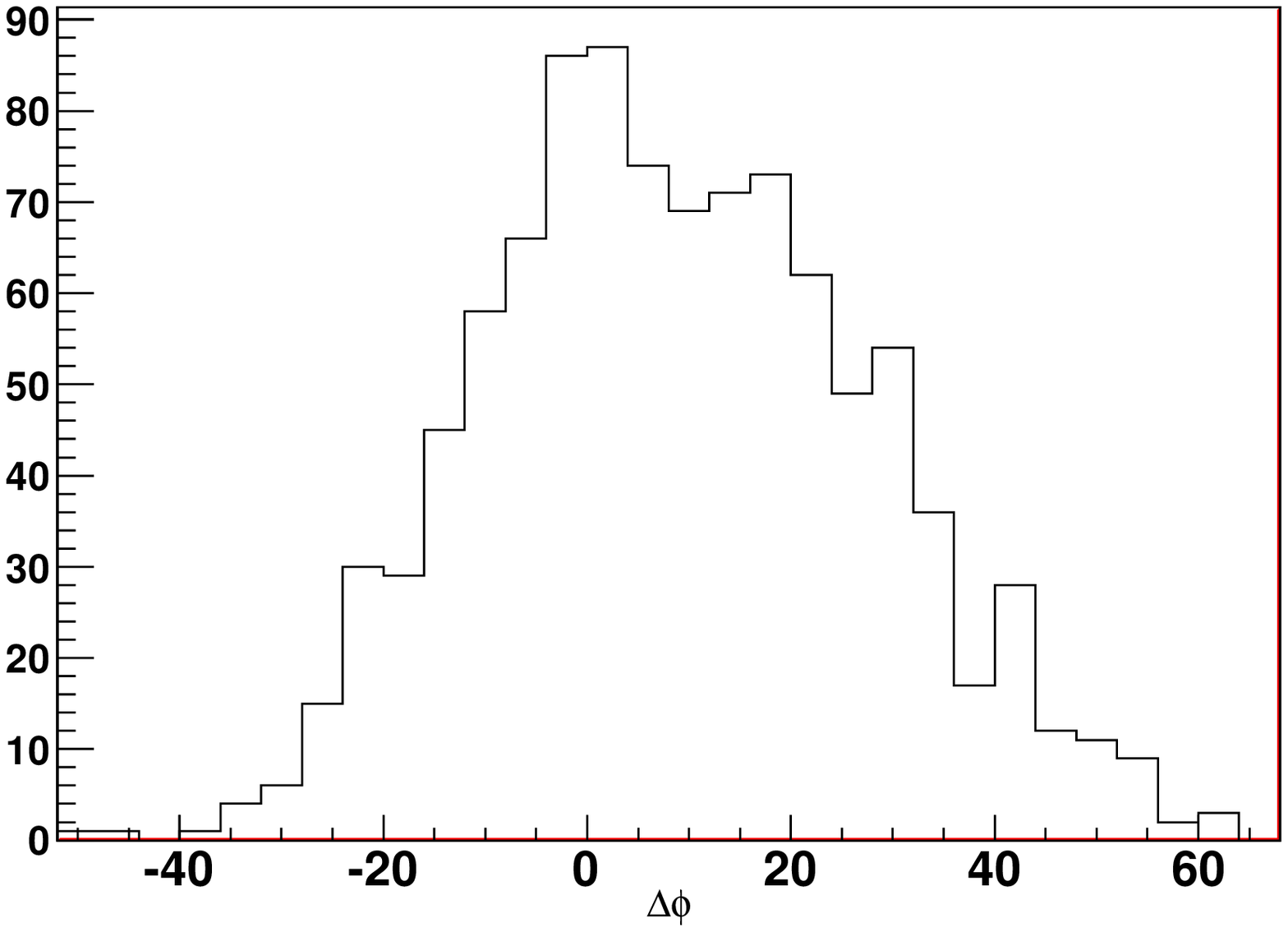}\\
\includegraphics[width=3.0in,angle=0]{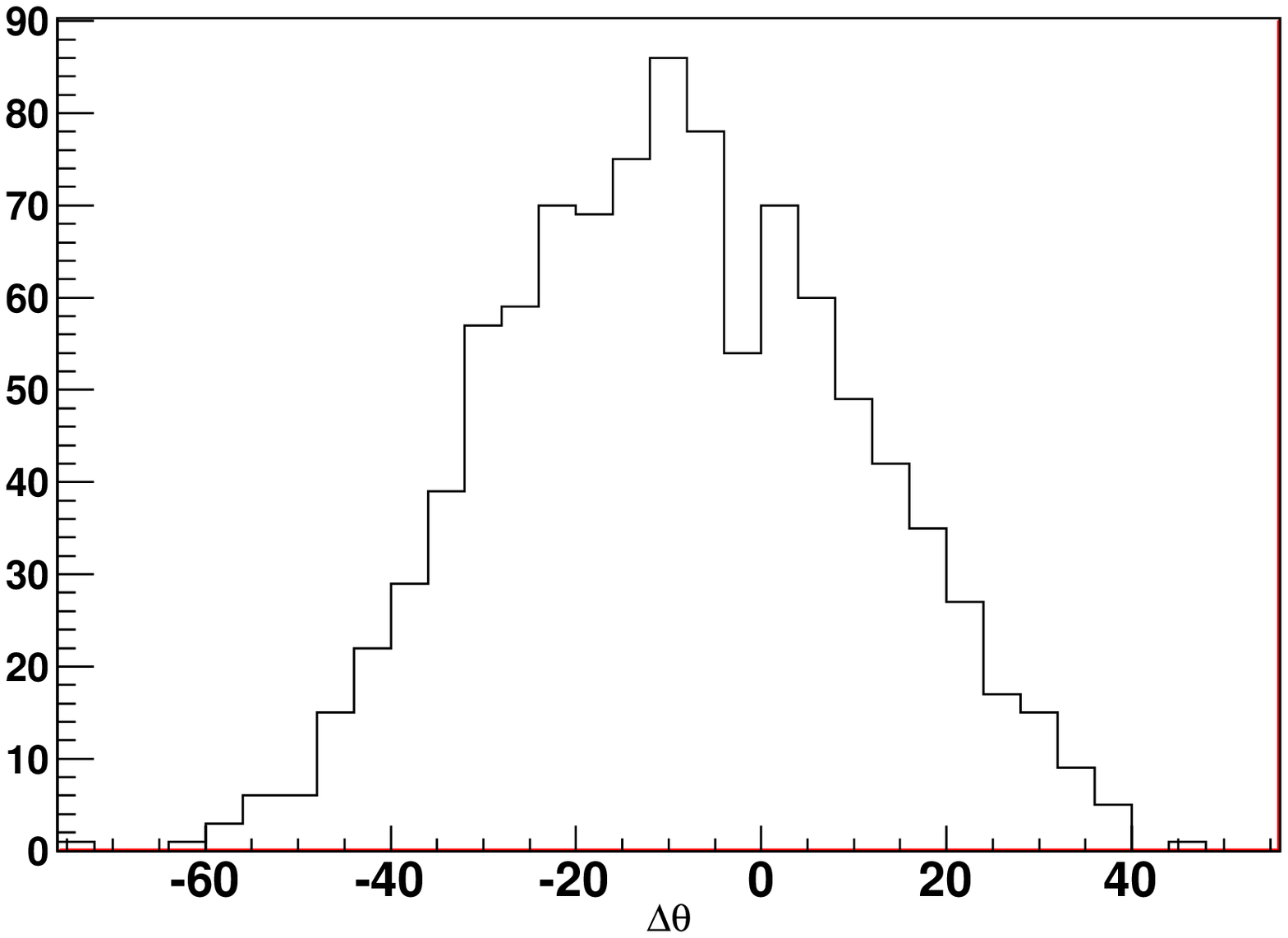}\\
\caption{
The
 distribution of bias 
parameters $k = \sqrt{C_1'/C_1}, \Delta \theta= \theta'-\theta$
 and $\Delta \phi = \phi'-\phi$ for source counts ($S>20$ mJy),
extracted by simulations. 
}
\label{fig:c1sim2_PI}
\end{figure}

The bias is computed by simulating a random sample of radio sources which 
has the same distribution as the real sources. 
The simulated sample has a dipole of approximately the same strength as seen in 
real NVSS data in roughly the same direction as observed.
 The pixels 
in the masked regions, $\delta \le -40^{\circ}$ and Galactic plane within the latitude $|b|<15^\circ$, are filled isotropically, as in the case of real data.  
This provides a particular random sample with same characteristics
as the real data.
 The results obtained from 1000 
realizations for sky brightness and number counts are shown in Fig. \ref{fig:c1sim_PI}, \ref{fig:c1sim2_PI} respectively. 
These figures show the distribution of $k$, 
$\Delta\theta$
and $\Delta \phi$. 
The resulting mean values of $k$, $\Delta\theta$
and $\Delta\phi$ over 1000 random samples represent their bias values.   
 {The extracted values of mean and standard deviation of
$k$, $\Delta \theta$ and $\Delta \phi$, both
for source count and sky brightness, are given in
 Table \ref{tb:bias} for different cuts on $S>S_{low}$. 
The mean values of $k$ are not too different from unity and the extracted
bias in angles is close to zero. 
Hence the required bias correction is relatively small. 
The standard deviations, given in Table \ref{tb:bias}, provide a 
reliable estimate of the statistical errors in the dipole parameters,
as discussed below.
}

\begin{table}[h!]
\begin{tabular}{|c|c|c|c|c|c|c|c|}
\hline
\multirow{2}{*}{$S_{\text{low}}$} & \multirow{2}{*}{Sources } & \multicolumn{3}{c|}{Source Counts} & \multicolumn{3}{c|}{Sky Brightness $S_I$}\tabularnewline
\cline{3-8}
 &  & $k$ & $\Delta\theta$ (deg) & $\Delta\phi$ (deg) & $k$ & $\Delta\theta$ (deg) & $\Delta\phi$ (deg)\tabularnewline
\hline
10 & 428210 & $0.85 \ (0.14)$ & $-8.9\ (16.0)$ & $  8.8\  (14.6)$ & $0.95\ (0.24)$ & $ -9.5\ (24.7)$ & $ 8.3\ (24.3)$\tabularnewline
\hline                                                                          
\hline                                                                          
20 & 240772 & $0.88\ (0.18)$ & $-9.2\ (19.5)$ & $  9.1\ (19.0)$ & $0.97\ (0.26)$ & $ -9.2\ (24.3)$ & $ 9.3\ (26.6)$\tabularnewline
\hline                                                                                                            
30 & 165206 & $0.91\ (0.21)$ & $-9.4\ (22.4)$ & $  9.7\ (22.0)$ & $1.01\ (0.28)$ & $ -8.8\ (27.5)$ & $ 9.6\ (29.4)$\tabularnewline
\hline                                                                                                            
40 & 124173 & $0.95\ (0.24)$ & $-9.9\ (24.2)$ & $ 10.3\ (24.9)$ & $1.06\ (0.32)$ & $ -9.3\ (28.6)$ & $ 9.3\ (31.6)$\tabularnewline
\hline                                                                                                            
50 & 98295 &  $0.99\ (0.27)$ & $-9.9\ (27.0)$ & $ 10.7\ (27.6)$ & $1.09\ (0.34)$ & $-10.6\ (30.4)$ & $12.6\ (32.9)$\tabularnewline
\hline
\end{tabular}
\caption{The values of $k$, $\Delta \theta$ and $\Delta\phi$ extracted from 
simulations corresponding to source counts and sky brightness, $S_I$. 
These values correspond to the bias generated in the dipole
amplitude and direction due to the filling of masked sky with randomly generated
data. The values in brackets are the standard deviations over 1000 samples.
}
\label{tb:bias}
\end{table}

The significance of dipole is calculated as follows. We generate 
10000 full sky isotropic random realizations of the NVSS data set. 
The random data
is generated by the procedure described above.  
We calculate the dipole power, $\tilde C_1$, of each of these random samples. 
The significance is equal to the probability that a 
random sample can generate a dipole of strength larger than that seen
in real data. We compute this by determining the number of random 
samples, whose dipole power exceeds that seen in real data. 
Here we use the data dipole power, $C'_1$, without applying bias correction
since it provides a conservative estimate of significance.  
The results
are presented by converting this  
probability into the sigma significance value. For example, two sigma
corresponds to a probability of 4.55\%.  

\subsection{Error Estimation}
{  
The bias simulations discussed above also give us a reliable estimate of 
statistical errors in dipole amplitude 
and angle variables--$\theta$ and $\phi$. These simulations take in account
all sources of random fluctuations, 
which include intrinsic fluctuations as well as those generated due 
to random filling of 
masked regions, that can affect the extracted dipole. Thus the standard 
deviation in the bias factor $k$, $\Delta\phi$ and $\Delta\theta$ directly gives an estimate of errors in dipole parameters. 
The fractional error in $k$ equals the fractional error in extracted dipole
amplitude. 
Additionally, the error in $\phi$ and $\theta$ is equal to $\Delta\phi$ and $\Delta \theta$ respectively.
}

\section{Results}
\label{sc:res}
\begin{figure}[!t]
\includegraphics[width=3.5in,angle=0]{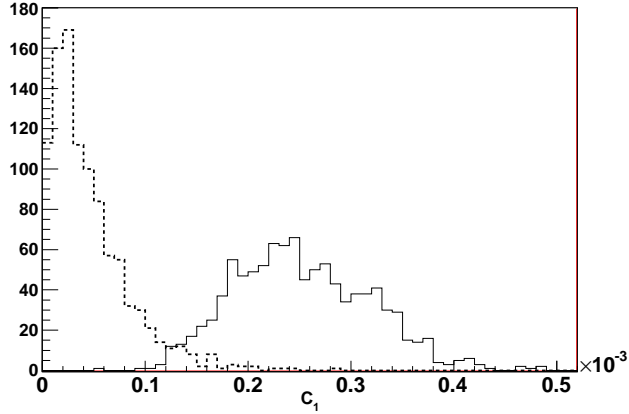}\\
\caption{ {The distribution of dipole power, $C'_1$, 
for number count of sources for the case
of real data (solid line) after imposing the cut 
$S>20$ mJy. The corresponding distribution of 
dipole power, $\tilde C_1$, 
 for random simulated data (dashed line) is also shown. 
The distribution of real data is obtained by randomly filling in the masked
regions, as explained in text. 
}}
\label{fig:c1count}
\end{figure}

\begin{figure}[!t]
\includegraphics[width=3.5in,angle=0]{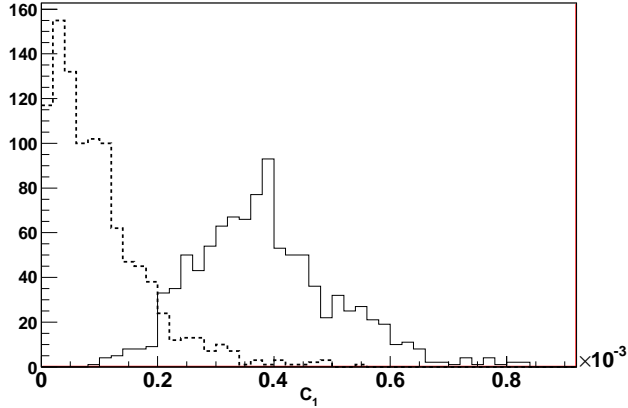}\\
\caption{ {The solid and dashed lines show the 
distributions of dipole power, $C'_1$,   
of real data and 
the dipole power, $\tilde C_1$, 
of random simulated data respectively for sky brightness. 
Here we impose the cut, $S>20$  mJy. 
The distribution of real data is obtained by randomly filling in the masked
regions, as explained in text.
}}
\label{fig:c1intensity}
\end{figure}

In Fig. \ref{fig:c1count}, we show the distribution of dipole power for
real data as well as randomly generated simulated data
 with the cut $S>20 $ mJy for the case of number counts. The corresponding
graphs for sky brightness, $S_I$, are shown in Fig. \ref{fig:c1intensity}.   
The sky brightness is obtained by summing the flux density 
over all sources in a particular
pixel.
 {The extracted values of $C'_1$ for various cuts 
 (S$>$10, 20, 30, 40, 50 mJy) 
for the case of
 number count and sky brightness, $S_I$, are given in 
Tables \ref{tb:resultCN}, \ref{tb:resultCI} for set (a).} 
In most cases we impose an upper limit $S<1000$ mJy on the flux density
of a source. We also test the sensitivity of our results to this upper limit. 
The cut $S>10$ mJy is not supposed to be reliable due to the bias 
in number counts \citep{Blake:2002}. We show it here mainly for
comparison to see how the results change as we push the limit on $S$ 
to lower values. Hence the results for this cut should be interpreted
cautiously. 
{The significance $(\sigma)$ of the detected dipole anisotropy as
well as the direction parameters $(Dec',RA')$ are also shown.
The primes on these parameters indicate that these have not been 
corrected for bias.}
 We find that the significance of the dipole anisotropy  
in number counts ranges from 3.2 $\sigma$ to 2.7 $\sigma$ for different
cuts. 
 For the sky brightness $(S_I)$ the maximum significance  
is found to be about 2.6 $\sigma$.

\begin{table}[h!]
\begin{tabular}{|c|c|c|c|c|c|}
\hline
\multirow{2}{*}{$S_{\text{low}}$} & \multirow{2}{*}{$C'_{1}\left(\times10^{4}\right)$} & \multirow{2}{*}{$\tilde{C}_{1}\left(\times10^{4}\right)$} & \multirow{2}{*}{significance} & \multirow{2}{*}{$DEC'$ (deg)} & \multirow{2}{*}{$RA'$ (deg)}\tabularnewline
 &  &  &  &  & \tabularnewline
\hline
10 & $1.28\pm0.35$ & $0.27\pm0.21$ & 3.0 $\sigma$ & $ 16\pm12$ & $146\pm 9$\tabularnewline
\hline
\hline
20 & $2.53\pm0.67$ & $0.46\pm0.38$ & 3.2 $\sigma$& $-13\pm11$ & $156\pm 9$\tabularnewline
\hline
30 & $3.10\pm0.86$ & $0.62\pm0.5$ & 3.0 $\sigma$& $-11\pm11$ & $159\pm 9$\tabularnewline
\hline
40 & $3.10\pm1.14$ & $0.83\pm0.63$ & 2.7 $\sigma$& $-28\pm12$ & $153\pm11$\tabularnewline
\hline
50 & $4.03\pm1.41$ & $1.03\pm0.80$ & 2.8 $\sigma$& $-29\pm11$ & $166\pm 9$\tabularnewline
\hline
\end{tabular}
\caption{The extracted value of the dipole power $C'_1$ and the corresponding
value for simulated isotropic data $\tilde C_1$ using number counts for 
different cuts on flux density of a source ($S>S_{low}$) for set (a). 
The significance of the dipole anisotropy, $\sigma$, as well as the
extracted polar angles of the dipole axis, 
$Dec'$ and $RA'$ are also shown. 
}
\label{tb:resultCN}
\end{table}

\begin{table}[h!]
\begin{centering}
\begin{tabular}{|c|c|c|c|c|c|}
\hline
\multirow{2}{*}{$S_{\text{low}}$} & \multirow{2}{*}{$C'_{1}\left(\times10^{4}\right)$} & \multirow{2}{*}{$\tilde{C}_{1}\left(\times10^{4}\right)$} & \multirow{2}{*}{significance} & \multirow{2}{*}{$Dec'$ (deg)} & \multirow{2}{*}{$RA'$ (deg)}\tabularnewline
 &  &  &  &  & \tabularnewline
\hline
10 &   $3.06\pm0.94$ &  $0.73\pm0.61$  &   2.7  $\sigma$ &   $ -4\pm14$  &   $162\pm 10$\tabularnewline
\hline
\hline
20 &   $3.84\pm1.22$ &  $0.98\pm0.83$  &   2.6 $\sigma$  &   $ -8\pm14$  &   $163\pm 10$\tabularnewline
\hline
30 &   $4.26\pm1.40$ &  $1.17\pm0.91$  &   2.6  $\sigma$ &   $ -7\pm15$  &   $163\pm 10$\tabularnewline
\hline
40 &   $4.32\pm1.58$ &  $1.34\pm1.10$  &   2.4  $\sigma$ &   $-11\pm16$  &   $162\pm 10$\tabularnewline
\hline
50 &   $4.86\pm1.75$ &  $1.46\pm1.16$  &   2.3 $\sigma$  &   $-11\pm16$  &   $165\pm 10$\tabularnewline
\hline
\end{tabular}
\par\end{centering}

\caption{The extracted value of the dipole power, $C'_1$, and the corresponding
value, $\tilde C_1$, for random isotropic data using sky brightness, $S_I$,
for different cuts $(S>S_{low})$ for set (a). 
The significance of the dipole anisotropy, $\sigma$, as well as the
extracted polar angles of the dipole axis, 
$DEC'$ and $RA'$ are also shown. 
}
\label{tb:resultCI}
\end{table}

In  
Tables \ref{tb:resultCN}, \ref{tb:resultCI}, we also list the mean dipole 
power, $\tilde C_1$, obtained from isotropic random samples. 
{It is clear that it is possible to detect a dipole anisotropy in real
data only if its power is significantly higher than $\tilde C_1$. For
comparison, the power expected due to kinematic dipole corresponding 
to $v=369$ Km/s is $C_1=2.97\times 10^{-5} $, assuming a pure power law
fit, Eq. \ref{eq:d2N}, with $x=1$. This is
smaller than the power corresponding to random isotropic samples for
most of the cases studied in Tables \ref{tb:resultCN}, \ref{tb:resultCI}. 
The only exception is the cut, $S_{low}=10$ mJy, for number counts. 
In this case also the random power is comparable to the power expected
due to CMBR dipole.
Hence
we find that it is not possible to extract a significant signal of
dipole anisotropy in NVSS data, if the only signal present in data arises 
due to the kinematic
effect, as expected from CMBR observations \cite{Crawford:2009}.
}

After correcting for bias, the extracted dipole amplitudes, 
 $|\vec{D}_N(obs)|$ and $|\vec{D}_S(obs)|$, corresponding to number counts
and sky brightness respectively, 
are shown in Tables \ref{tb:resultDN}, \ref{tb:resultDI}. 
The extracted speed of the solar system, assuming that dipole is entirely
a kinematic effect, and angles
 $(Dec,RA)$, after correcting for bias, are also shown. 
 {The speed is extracted by using the improved fit, 
Eq. \ref{eq:d2Ng}. 
The values of $x,\beta$ in Eq. \ref{eq:d2Ng},  
 obtained by directly fitting the data sample, are given in Table
\ref{tb:improved_para}. Using the pure power law fit, we find that the
 extracted speed is  similar to that obtained by the
improved fit. }
{In Tables \ref{tb:resultDN}, \ref{tb:resultDI} we give results for 
both the data sets (a) and (b). The data sets are described in 
section \ref{sc:data}. We point out that set (b) is obtained after 
imposing a more stringent cut on the data for removal of local sources. 
We find that the extracted
dipole amplitude for set (b) is smaller in comparison to
set (a) by about 20\% for the case of number counts. However the
change is much smaller for sky brightness. In most cases, the final result
is still more than three times the speed expected from CMBR dipole.  
The shift in the
direction parameters from set (a) to set (b) is found to be well within
errors.
}  

We find that, for the case of number counts, the results show some
variation with the cut on flux density. However the change is not
very large and the results agree within errors,
as long as we ignore the cut $S>10$ mJy. For such small values of
flux density, the data set is known to have uneven distribution of source
counts due to non-uniform sampling \citep{Blake:2002}. The corresponding
parameters, extracted using sky brightness, are found to be comparatively
insensitive to the cut imposed. {The angle parameters show almost 
no change, whereas the extracted speed varies between $1140\pm 470$ 
to $1210\pm460$, including the cut $S>10$ mJy. 
The extracted speed is found to be about $3$ times the expectation from 
CMBR observations. The significance of the difference is about $3$ $\sigma$
for set (a), both for number counts and sky brightness.
For set (b) the significance of difference is roughly 2 $\sigma$.  
Hence the data sample is not consistent with the CMBR dipole.
However the deviation is found to be less significant in comparison
to earlier results 
 \citep{Singal:2011,Gibelyou:2012,Rubart:2013}. 
Our result 
indicates the presence of an intrinsic dipole anisotropy which
cannot be explained in terms of local motion. }

We next test the sensitivity of our results to the choice of 
pixel size. Using the HEALPix resolution parameter, $N_{side}=16$,
we find that results for dipole parameters agree
with those obtained with $N_{side}=32$ within
errors. 
In particular, the results for brightness
show very little sensitivity to the choice of pixel size.
The extracted dipole power for number counts, however, 
is found to be systematically
larger by 3\% to 7\% for different cuts on $S$. 
We have also tested how the results change if we
increase the exclusion radius around each masked local source from 30 arcsecs
to 45 arcsecs. The results for number counts change by less than 1 percent for
all the cuts considered. For sky brightness, the change is less than 3 percent
for all cases. Hence we find that the extracted dipole parameters are not very
sensitive to the choice of exclusion radius.

\begin{table}[h!]
\begin{tabular}{|c|c|c|c|c|c|}
\hline
\multirow{2}{*}{$S_{\text{low}}$}  
& \multirow{2}{*}{Set}
& \multirow{2}{*}{$|\vec{D}_{N}|$} 
& \multicolumn{3}{c|}{$\vec{v}$}\tabularnewline
\cline{4-6}
 & &  & $|\vec{v}|$ (Km/s)& RA (deg) & DEC (deg)\tabularnewline
\hline
10 & (a) & $0.0113\pm0.0018$  & $1020\pm170$  & $137\pm 15$ & $  8\pm16$\tabularnewline
 & (b) & $0.0096\pm0.0026$  & $810\pm220$  & $145\pm 20$ & $  20\pm17$\tabularnewline
\hline                                                          
\hline                                                          
20 &(a) & $0.0153\pm0.0032$ & $1290\pm270$ & $147\pm 19$ & $-22\pm19$\tabularnewline
 &(b) & $0.0125\pm0.0040$ & $1000\pm320$ & $159\pm 27$ & $-15\pm22$\tabularnewline
\hline                                                          
30 & (a)&$0.0163\pm0.0038$ & $1320\pm310$ & $149\pm 22$ & $-21\pm22$\tabularnewline
 & (b)&$0.0143\pm0.0048$ & $1110\pm370$ & $159\pm 34$ & $-14\pm25$\tabularnewline
\hline                                                          
40 &(a) &$0.0157\pm0.0040$ & $1230\pm310$ & $143\pm25$  & $-38\pm24$\tabularnewline
 &(b) &$0.0136\pm0.0049$ & $1030\pm370$ & $156\pm38$  & $-31\pm27$\tabularnewline
\hline                                                          
50 & (a)&$0.0172\pm0.0047$  &$1320\pm360$ & $156\pm28$  & $-39\pm27$\tabularnewline
 &(b) &$0.0157\pm0.0059$  &$1160\pm440$ & $175\pm43$  & $-33\pm28$\tabularnewline
\hline
\end{tabular}
\caption{The extracted dipole amplitude $|\vec{D}_{N}(obs)|$ for different cuts,
after correcting for bias. The corresponding parameters of the velocity
vector, $\vec v$, of the solar system are also shown. 
 {We give results both for data sets (a) and (b). Here set (b) is
is obtained by imposing a 
more stringent cut, as explained in text. }
}
\label{tb:resultDN}
\end{table}

\begin{table}[h!]
\begin{tabular}{|c|c|c|c|c|c|}
\hline
\multirow{2}{*}{$S_{\text{low}}$} 
& \multirow{2}{*}{Set}
& \multirow{2}{*}{$|\vec{D}_{S}|$}
 & \multicolumn{3}{c|}{$\vec{v}$}\tabularnewline
\cline{4-6}
 & & & $|\vec{v}|$ (Km/s)& RA (deg)& DEC (deg)\tabularnewline
\hline
10 & (a) &$0.0155\pm0.0040$&  $1220\pm310$ & $153\pm24$ & $-13\pm25$\tabularnewline
 & (b) &$0.0141\pm0.0052$&  $1190\pm440$ & $166\pm39$ & $-8\pm26$\tabularnewline
\hline                                                         
\hline                                                         
20 &(a) &$0.0171\pm0.0045$  & $1300\pm340$ & $153\pm27$ & $-17\pm24$\tabularnewline
&(b) &$0.0151\pm0.0057$  & $1210\pm460$ & $168\pm43$ & $-13\pm30$\tabularnewline
\hline                                                         
30 &(a) &$0.0172\pm0.0048$  &$1280\pm360$ & $153\pm29$ & $-16\pm28$\tabularnewline
 &(b) &$0.0153\pm0.0059$  &$1180\pm450$ & $168\pm46$ & $-11\pm30$\tabularnewline
\hline                                                         
40 &(a) &$0.0165\pm0.0050$ &  $1210\pm370$ & $153\pm32$ & $-20\pm29$\tabularnewline
 &(b) &$0.0151\pm0.0063$ &  $1140\pm470$ & $172\pm51$ & $-15\pm32$\tabularnewline
\hline                                                         
50 &(a) &$0.0171\pm0.0053$ &$1240\pm380$ & $152\pm33$ & $-21\pm30$\tabularnewline
 &(b) &$0.0160\pm0.0065$ &$1190\pm470$ & $174\pm50$ & $-15\pm32$\tabularnewline
\hline
\end{tabular}
\caption{The dipole amplitude, $|\vec{D}_{S}(obs)|$,
extracted from sky brightness, $S_I$, for different cuts,
after correcting for bias. 
{We give results both for set (a) and (b).
} 
}
\label{tb:resultDI}
\end{table}

The results obtained by using sky brightness are found to be relatively 
insensitive to the lower limit imposed on the flux density. However it is
possible that these results might be sensitive to the upper limit, 
especially since the integral of $d^2S_{I,obs}$, Eq. \ref{eq:d2S1}, using
a pure power law fit, \ref{eq:d2N}, diverges if the
limits of integration are taken to be $S_{low}$ to $\infty$. Hence it is
important to test the sensitivity of our results to the upper limit
also in this case. The results obtained using an upper limit $S_{up}=900$ mJy,
instead of $S_{up}=1000$ mJy used earlier, show negligible change. 
The dipole amplitude for this case is found to be, 
$0.0149\pm0.0036$,   
 $0.0162\pm0.0045$,   
 $0.0164\pm0.0046$,  
$0.0161\pm0.0049$,   
 $0.0167\pm0.0053$,  
for the cuts, $S>10, 20, 30, 40, 50$ mJy respectively for set (a).
We find that the dipole
amplitude changes by only about 5 percent in comparison to the
results shown in Table \ref{tb:resultDI}. The direction parameters show 
even smaller change. 
Hence we do not find 
strong sensitivity to either the upper limit or the lower limit in 
this case.

\begin{figure}[!t]
\includegraphics[width=4.5in,angle=0]{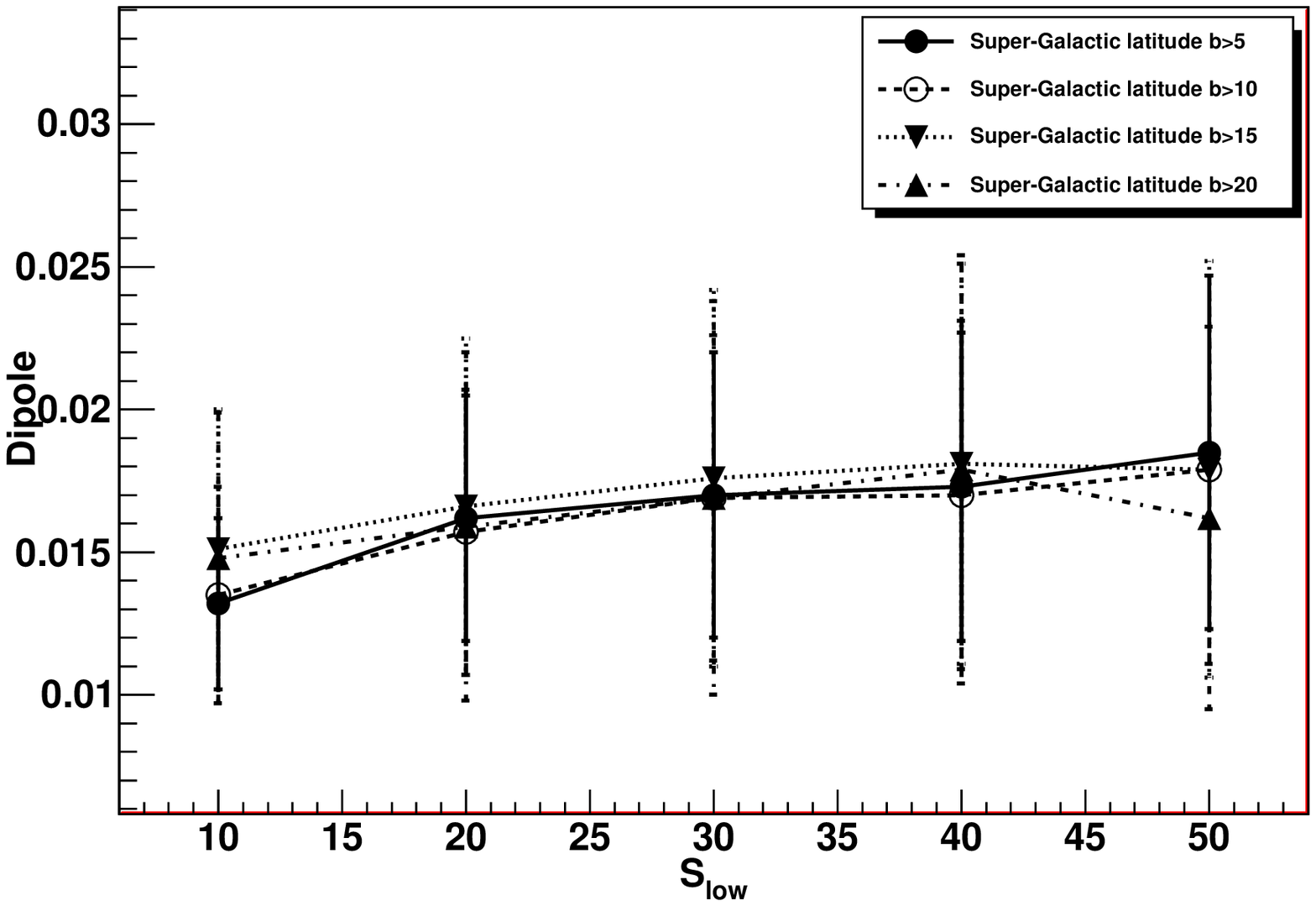}\\
\includegraphics[width=4.5in,angle=0]{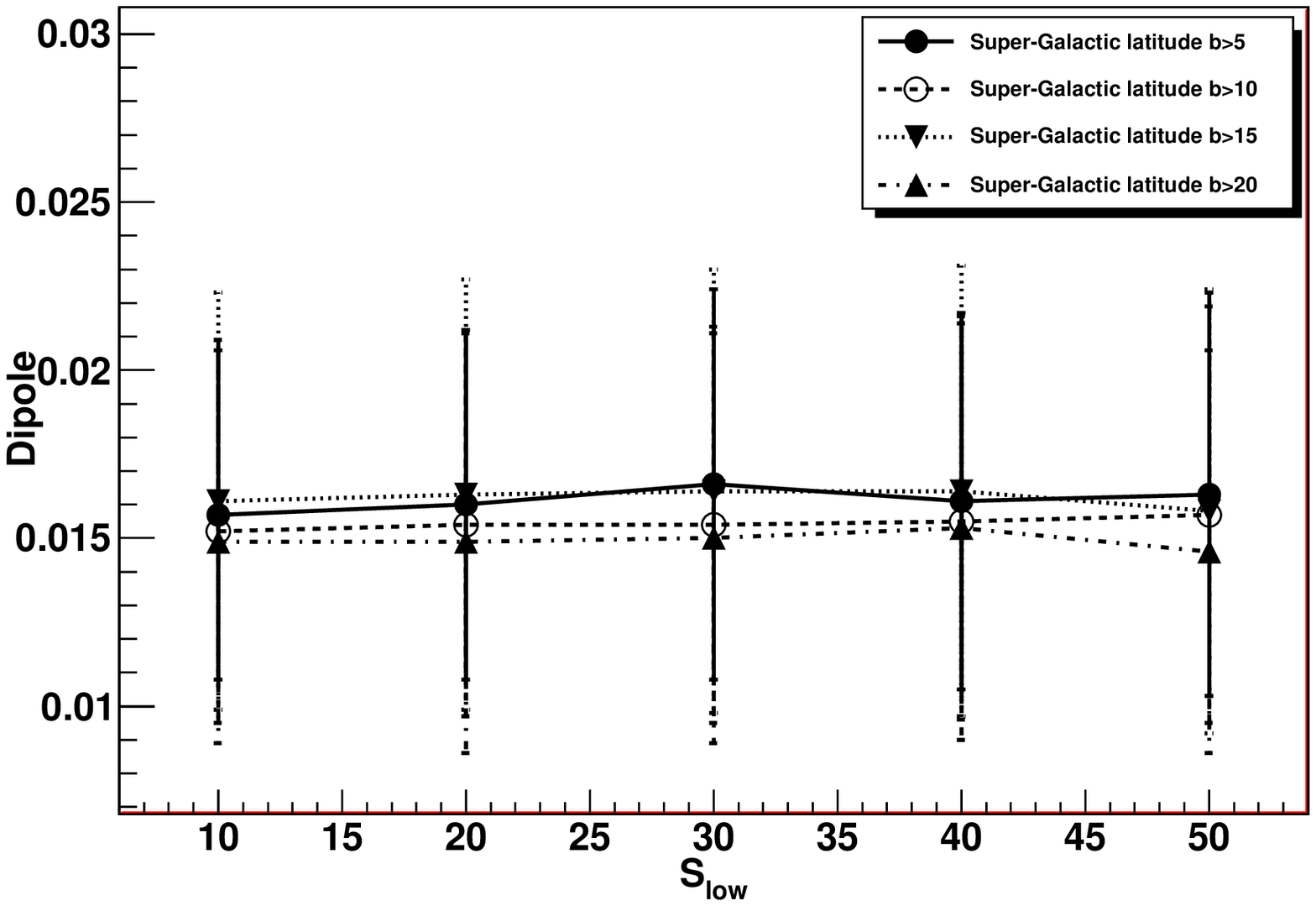}\\
\caption{{The dipole amplitude obtained after imposing a
cut on the supergalactic latitude, $|b'|<5^o,10^o,15^o,20^o$.  
The upper and lower graphs correspond to number counts and
sky brightness respectively.
}}
\label{fig:sens_supgcut}
\end{figure}

{An alternative procedure for the removal of local sources is to
mask the supergalactic plane. We expect that the local sources would
be dominantly clustered close to this plane. Hence their contribution
would be significantly reduced if this plane is masked out. In 
Fig. \ref{fig:sens_supgcut} we show the dipole amplitude for different
cuts on the flux density, $S$, after masking the supergalactic plane. 
In this case we do not remove the local sources using known catalogues.
 We expect that their effect would be minimized 
by the cut on supergalactic plane. 
The results are shown after masking regions corresponding to 
supergalactic latitude, $|b'|<5^o, 10^o, 15^o$ and $20^o$. In this
analysis the galactic plane, corresponding to $|b|<15^o$, is also masked. 
We find that the results obtained are consistent with those shown
in Tables 
\ref{tb:resultDN}
and \ref{tb:resultDI}. The dipole, both for the number counts and 
sky brightness, does not show
any particular trend. 
The overall dependence on the cut on $|b'|$ is well
within errors on the dipole amplitude. 
We point out that 
the masked region for the cut $|b'|<15^o, 20^o$ becomes very large. 
For example, for the cut, $|b'|<20^o$,
we find that 60\% of the sky gets eliminated. In comparison, for 
$|b'|<10^o$, 49\% gets eliminated.
For large masks, our
procedure of bias correction may not be very reliable. Hence some of these
results, particularly for cuts, $|b'|<15^o, 20^o$, 
should be interpreted with caution.  
In any case it is satisfying that the results using this approach are 
comparable to those obtained in Tables \ref{tb:resultDN} and \ref{tb:resultDI}
 by directly removing local sources using standard catalogues.
}

\begin{figure}[!t]
\includegraphics[width=4.5in,angle=0]{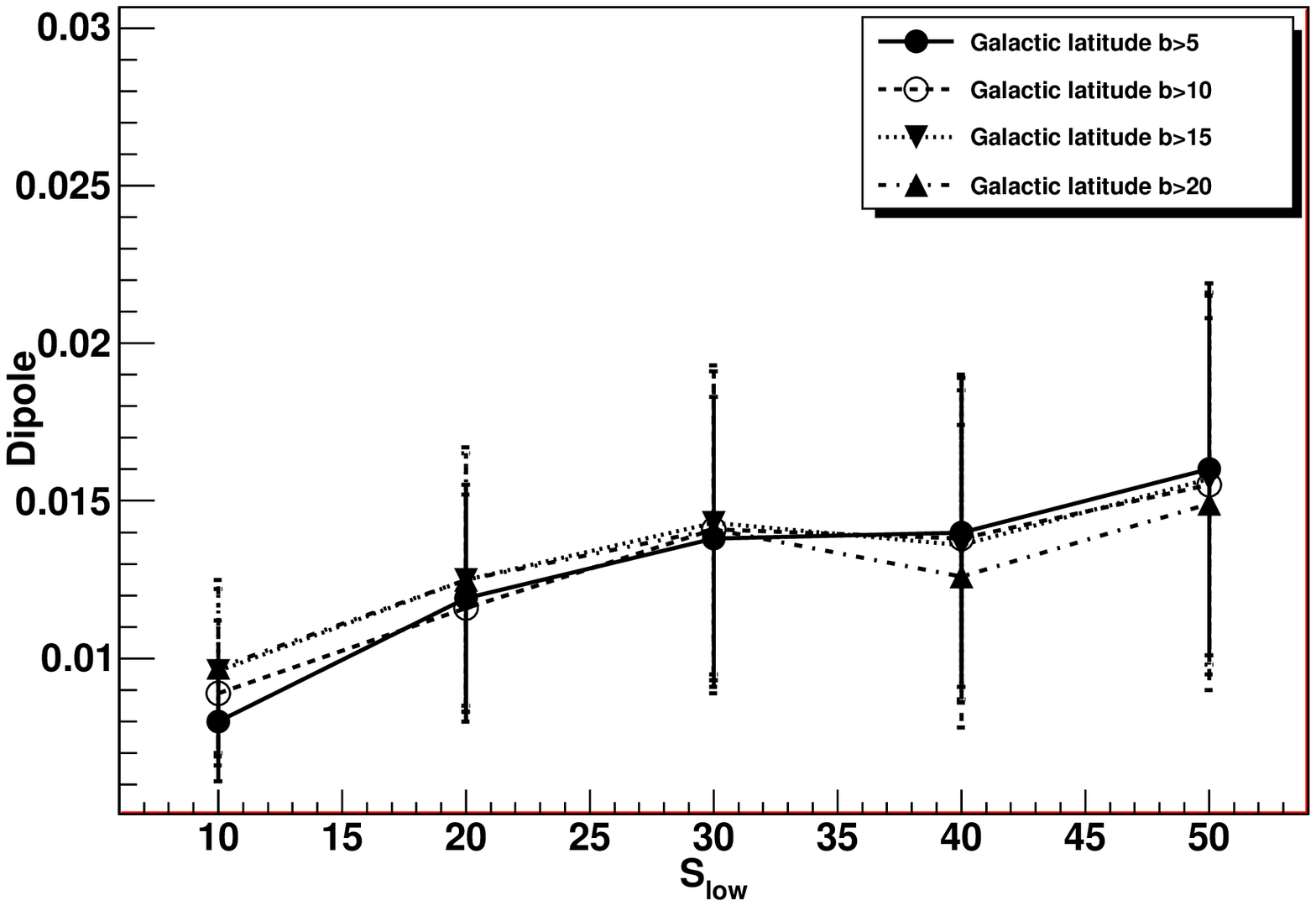}\\
\includegraphics[width=4.5in,angle=0]{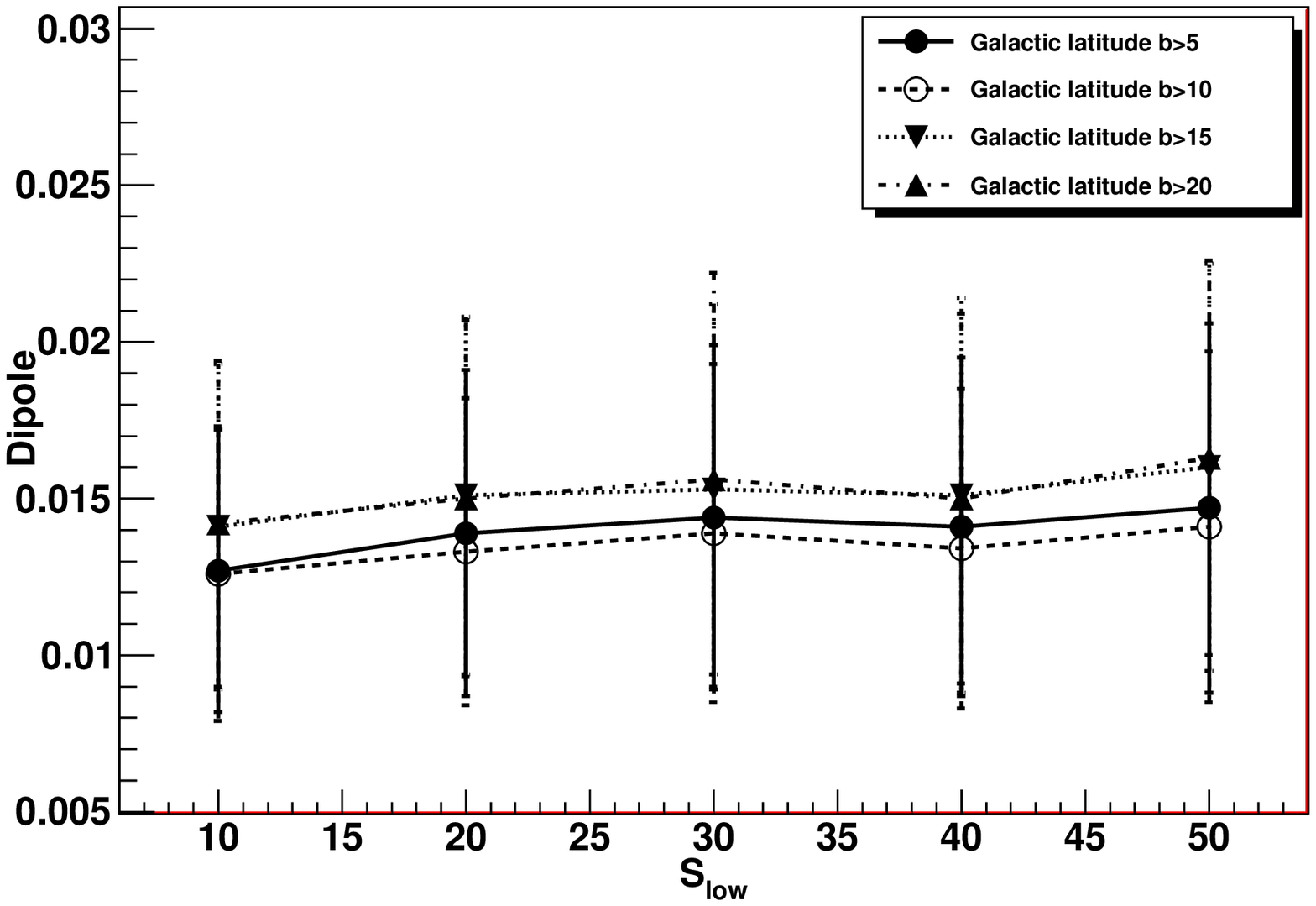}\\
\caption{ {The sensitivity of the extracted dipole amplitude
to the cut imposed to remove the galactic plane. Here we show results
after removing regions corresponding to galactic latitude,
$|b|<5^o,10^o,15^o,20^o$.  
The upper and lower graphs correspond to number counts and
sky brightness respectively.
}}
\label{fig:sens_gcut}
\end{figure}

 {We next determine the sensitivity of our results to the cut on the
galactic plane. In this case we use data set (b), i.e. we impose
a stringent cut to remove the local sources.  
In Fig. \ref{fig:sens_gcut}, we show the change in dipole amplitude
as we mask the galactic plane corresponding to latitude, $|b| <5^o, 10^o,
15^o$ and $20^o$. We again do not find any particular trend in the
dipole amplitude, both for number counts and sky brightness, 
as we increase $|b|$ from $5^o$ to $20^o$. 
For all the five cuts on $S$ the dependence on the
cut on galactic latitude is relatively mild.
We also consider a galactic cut which depends on the galactic longitude
$l$ in order to remove a larger region near the galactic center. In this
case we exclude the region corresponding to, $|b| <25^o$  for $|l|<20^o$,
and  $|b|<10^o$ for other values of $l$.
In this case we find that the dipole amplitude is $0.0083, 
0.0115, 0.0138, 0.0134$ and $0.0144$ for the cuts $S>10, 20, 30, 40, 50$
mJy respectively 
for the case of number counts. The corresponding dipole amplitude
for sky brightness is $0.0127, 0.0138, 0.0144, 0.0139, 0.0141$. Again 
by comparing with Fig. \ref{fig:sens_gcut}, we find results consistent
with other cuts used to eliminate galactic plane. 
 Hence we conclude that the
extracted dipole is not very sensitive to this cut.  
}

\subsection{Dipole in flux density per source}
Finally, we briefly discuss the results obtained by using the measure
flux density per source, $\bar S$. 
For the case of pure power law fit, Eq. \ref{eq:d2N}, 
 the kinematic dipole is absent
in this parameter. 
The improved fit, Eq. \ref{eq:d2Ng}, does lead to a non-zero kinematic
dipole but its amplitude is relatively small.  
The extracted dipole power for data and isotropic random samples is
given in Table \ref{tb:resultCNbarS}. 
In this case the dipole anisotropy 
is not found to be significant for any the cuts, $S>10,20,30,40,50$ mJy. 
This is consistent with our expectations. Furthermore,
this result supports our assumption that any intrinsic dipole, which may
be present in data, dominantly affects the number counts. 
This method may 
be used more effectively with a larger data set and might provide an 
independent probe of our local motion. As explained in section 2.1, it might
allow an independent extraction of both the local velocity and the intrinsic
dipole in number counts.

\begin{table}[h!]
\begin{tabular}{|c|c|c|c|}
\hline
\multirow{2}{*}{$S_{\text{low}}$} & \multirow{2}{*}{$C'_{1}\left(\times10^{4}\right)$} & \multirow{2}{*}{$\tilde{C}_{1}\left(\times10^{4}\right)$} & \multirow{2}{*}{significance}  \tabularnewline
 &  &  &   \tabularnewline
\hline
10 & $1.00\pm0.51$ & $0.50\pm0.39$ & $1.6 \sigma$ \tabularnewline
\hline
\hline
20 & $0.43\pm0.29$ & $0.58\pm0.49$ & $0.6 \sigma$ \tabularnewline
\hline
30 & $0.40\pm0.33$ & $0.64\pm0.53$ & $0.5 \sigma$ \tabularnewline
\hline
40 & $0.61\pm0.40$ & $0.68\pm0.56$ & $0.8 \sigma$ \tabularnewline
\hline
50 & $0.56\pm0.43$ & $0.72\pm0.59$ & $0.6 \sigma$ \tabularnewline
\hline
\end{tabular}
\caption{The extracted value of the dipole power $C'_1$ and the corresponding
value for simulated isotropic data $\tilde C_1$ using flux density per 
source, $\bar S$ for 
different cuts. The corresponding 
 significance of the dipole anisotropy, $\sigma$, is also given.
As expected, we find that the dipole anisotropy is not significant 
in this case.
}
\label{tb:resultCNbarS}
\end{table}

\section{ Discussion and Conclusion}
\label{sc:physics}
We qualitatively confirm the results obtained in \citep{Singal:2011}.
We find that the dipole anisotropy, both in number counts and sky brightness,
cannot be consistently interpreted in terms of the local motion of 
the solar system, as derived by the CMBR measurements.  {The difference
is significant at roughly $2$ $\sigma$, both for number counts and sky
brightness.}
 The results for the case of sky 
brightness are relatively insensitive to the cut imposed.  
We also test the sensitivity of our results for this case  
 to the upper limit imposed on the flux density. We find that
the extracted dipole is relatively insensitive to the upper limit.   
Our extracted speed, using spherical harmonic decomposition, 
is somewhat smaller in comparison to
\citep{Singal:2011}.
The difference probably arises due to the 
procedure used in extracting the dipole. We make a spherical harmonic
decomposition of data, which isolates the dipole contribution. 
In contrast the  
procedure used in \citep{Singal:2011} would also get contributions from 
higher multipoles, which may be small but not completely negligible. 
{We also impose a more stringent cut to remove local sources 
in comparison to that used in  
\citep{Singal:2011}.
In particular, we remove bright and extended sources,
which are misidentified in the NVSS survey as a large number of sources,
and hence introduce spurious correlations. Furthermore we use the
more extensive catalogues which have recently become available to remove
local sources, which contribute to the clustering dipole.  
We find that these more stringent cuts lead to a somewhat lower 
amplitude of the extracted dipole.
}

We extract the local speed by using an improved fit to the number 
density, $n(S)$, as a function of the flux density $S$. 
This fit takes into account the deviation of $n(S)$ from a pure power law.
 {We find that the results obtained by this improved fit are comparable 
to that obtained by a pure power law fit.} 
We also find that this method leads to 
slightly different 
kinematic dipole in number counts in comparison to sky brightness. 
We argue that,
in principle, this can be used to independently extract the local speed
and an intrinsic dipole, which may be present in data. 
This procedure may be used more effectively when a larger data set becomes
available.
This fit also allows an independent extraction of local speed
using flux density per source. With the present data set, however, 
this does not lead to a significant extraction of dipole anisotropy, 
which is expected to be small in this case. 

We conclude that the dipole amplitude in NVSS data is significantly
larger in comparison to the prediction based on CMB dipole. The direction,
however, is close to the CMB dipole.



\section*{Acknowledgements}
We have used CERN ROOT 5.27 for generating our plots. 
Some of the results in this paper have been derived using the HEALPix 
\citep{Gorski:2004by} package. 
Prabhakar Tiwari and Rahul Kothari 
sincerely acknowledge CSIR, New Delhi for the award of fellowship during the work. We thank Ashok K. Singal and Chris Blake for useful comments on our paper.  
We also thank Chris Blake for providing details about the cuts imposed
in \citep{Blake:2002}.
\bibliographystyle{elsarticle-num}
\bibliography{radioNVSS}
\end{document}